\documentclass[fleqn,usenatbib]{mnras}
\usepackage{newtxtext,newtxmath}
\usepackage{booktabs}
\usepackage{graphicx}
\usepackage{hyperref}
\usepackage[normalem]{ulem}
\usepackage{soul}
\usepackage[T1]{fontenc}
\usepackage{ae,aecompl}
\usepackage{url}

\usepackage{graphicx}	
\usepackage{amsmath}	
\usepackage{amssymb}	
\usepackage{float}
\usepackage{array}
\newcommand{\PreserveBackslash}[1]{\let\temp=\\#1\let\\=\temp}
\newcolumntype{C}[1]{>{\PreserveBackslash\centering}p{#1}}
\newcolumntype{R}[1]{>{\PreserveBackslash\raggedleft}p{#1}}
\newcolumntype{L}[1]{>{\PreserveBackslash\raggedright}p{#1}}
\usepackage{caption}
\newcolumntype{H}{>{\setbox0=\hbox\bgroup}c<{\egroup}@{}}
\newcommand\setrow[1]{\gdef\rowmac{#1}#1\ignorespaces}
\newcommand\clearrow{\global\let\rowmac\relax}
\clearrow

\newcommand{\gr}{$\gamma$-ray}
\newcommand{\nup}{$\nu_{\rm peak}^S$}
\newcommand{\fermi}{{\it Fermi}}

\newcommand{\swift}{{\it Swift}}

\usepackage{lineno}
\newcommand{\lsim}{{\lower.5ex\hbox{$\; \buildrel < \over \sim \;$}}}
\newcommand{\gsim}{{\lower.5ex\hbox{$\; \buildrel > \over \sim \;$}}}

\title[Dissecting the regions around IceCube high-energy neutrinos]{Dissecting the regions around IceCube high-energy neutrinos: growing evidence for the blazar connection}

\author[P. Giommi et al.]{P. Giommi$^{1,2,3}$, T. Glauch$^{1, 4}$, P. Padovani$^{5,6}$, E. Resconi$^4$, A. Turcati$^4$, Y.L. Chang$^{3,7}$\\ 
$^{1}$Institute for Advanced Study, Technische
Universit{\"a}t M{\"u}nchen, Lichtenbergstrasse 2a,
D-85748 Garching bei M\"unchen, Germany\\
$^{2}$Associated to Agenzia Spaziale Italiana, ASI, via del Politecnico s.n.c., I-00133 Roma Italy \\
$^{3}$ICRANet, Piazzale della Repubblica,10 - 65122, Pescara,
Italy\\
$^{4}$Technische Universit{\"a}t M{\"u}nchen, Physik-Department,
James-Frank-Str. 1, D-85748 Garching bei M{\"u}nchen, Germany\\
$^{5}$European Southern Observatory, Karl-Schwarzschild-Str.
2, D-85748 Garching bei M\"unchen, Germany\\
$^{6}$Associated to INAF - Osservatorio Astronomico di Roma, via Frascati 33,
I-00040 Monteporzio Catone, Italy\\
$^{7}$Tsung-Dao Lee Institute, Shanghai Jiao Tong University, 800 Dongchuan RD. Minhang District, Shanghai, China\\
}

\date{Accepted XXX. Received YYY; in original form ZZZ}

\pubyear{2020}

\begin{document}
\label{firstpage}
\pagerange{\pageref{firstpage}--\pageref{lastpage}}
\maketitle

\begin{abstract}
The association of two IceCube detections, the IceCube-170922A event and a neutrino flare, with the blazar TXS\,0506+056, has paved the way for the multimessenger quest for cosmic accelerators. IceCube has observed many other neutrinos but their origin remains unknown. To better understand the reason for the apparent lack of neutrino counterparts we have extended the comprehensive dissection of the sky area performed for the IceCube-170922A event to all 70 public IceCube high-energy neutrinos that are well reconstructed and off the Galactic plane.
Using the multi-frequency data available through the Open Universe platform, we have identified numerous candidate counterparts of IceCube events. We report here the classification of all the \gr\, blazars found and the results of subsequent statistical tests. In addition, we have checked the 4LAC, 3FHL and 3HSP catalogues for potential counterparts.
Following the dissection of all areas associated with IceCube neutrinos, we evaluate the data using a likelihood-ratio test and find a $3.23\,\sigma$ (post-trial) excess of HBLs and IBLs with a best-fit of $15 \pm 3.6$ signal sources. This result, together with previous findings, consistently points to a growing evidence for a connection between IceCube neutrinos and blazars, the most energetic particle accelerators known in the Universe.
\end{abstract}



\section{Introduction}\label{sec:Introduction}

The IceCube Neutrino Observatory at the South Pole\footnote{\url{http://icecube.wisc.edu}} has reported in the past few years on the detection of tens of high-energy neutrinos of likely astrophysical origin \citep{2013PhRvL.111b1103A,Aartsen2015,Aartsen2016,ICECube13,ICECube14,ICECube15_1,ICECube17_1,ICECube17_2,2019ICRC...36.1004S,2019ICRC...36.1017S}. 
This result, together with the recent association in space and time between the 
bright blazar TXS\,0506+056 and one IceCube neutrino detected in Sept. 
2017 and a neutrino flare in 2014-2015  \citep{icfermi,iconly,dissecting}, is triggering a 
large interest in the nature of the electromagnetic counterparts of astrophysical neutrinos.

High-energy neutrinos in a cosmic context are thought to be generated when very high-energy 
(VHE) cosmic rays interact with matter or radiation creating charged and neutral mesons, 
which then decay into neutrinos, $\gamma$-rays and other particles. Neutrinos and 
$\gamma$-rays are the ``messengers'' that can travel cosmological distances and reach the 
Earth undeflected, thus providing information on the VHE physical processes that generated 
them.

Blazars, the most abundant type of \gr\ sources in the extra-galactic sky \citep[e.g.][]{Fermi3LAC}, have 
long been suspected to be capable of accelerating cosmic rays to sufficiently large energies to produce 
astrophysical neutrinos \citep[e.g.][]{mannheim95,halzen97,mueckeetal03}. Blazars are a rare type of Active 
Galactic Nuclei \citep[AGN; see][for a review]{Padovani_2017} characterised by the emission of strong and 
highly variable non-thermal radiation across the entire electromagnetic spectrum. This radiation is 
generated by energetic charged particles moving in a magnetic field within a relativistic jet that is seen 
at a small angle with respect to the line of sight \citep{UP95,Padovani_2017}.

Blazars come in two flavours depending on the presence and on the strength of their optical emission lines, 
namely Flat Spectrum Radio Quasars (FSRQs) when
their optical spectrum shows broad emission lines just like standard broad-lined AGN, and BL Lacs when their optical 
spectrum is featureless or it includes weak emission lines with equivalent width $<
5$ \AA~\citep{Stickel_1991,Stocke_1991}. From a spectral energy distribution (SED) point 
of view blazars can further be divided into low, intermediate, and high energy peaked 
objects (LBLs, IBLs, and HBLs respectively) depending on the energy where the power of 
their synchrotron emission peaks (\nup\footnote{LBL: \nup~$<10^{14}$~Hz; 
IBL: $10^{14}$~Hz$<$ \nup~$<10^{15}$~Hz; HBL: \nup~$>10^{15}$~Hz.}) in their 
SED \citep[][]{padgio95}.

Several studies have reported hints of a correlation between blazars and 
the arrival direction of astrophysical neutrinos 
(\citealt{Pad_2014,Padovani_2016,Lucarelli_2017,Lucarelli_2019}; but see also \citealt{ICECube17_3}) and possibly
of Ultra High Energy Cosmic Rays \citep{Resconi_2017}.

\cite{Padovani_2016} have correlated the second catalogue of hard {\it Fermi}-LAT sources 
(2FHL, $E>50$\,GeV, \citealt{2FHL}) and other catalogues, with the then publicly available 
high-energy neutrino sample detected by IceCube. The chance probability of association of 
2FHL HBLs with IceCube events was 0.4 per cent, which becomes 1.4 per cent ($2.2\,\sigma$) 
by evaluating the impact of trials \citep{Resconi_2017}. This hint appears to be strongly dependent 
on $\gamma$-ray flux. The corresponding fraction of the IceCube signal explained by HBLs is however only 
$\sim 10 - 20$ per cent, which agrees with the results of \cite{Aartsen2017,ICECube17_3,Huber:2019lrm}, who 
by searching for cumulative neutrino emission from blazars in the second {\it Fermi}-LAT 
AGN \citep[2LAC;][]{Fermi2LAC} and other catalogues (including also the 2FHL and 3FHL), have constrained 
the maximum contribution of known blazars to the observed astrophysical neutrino flux to $< 17 - 27$ 
per cent.

All the neutrino events that correlate with a source in  \cite{Padovani_2016} and \cite{Resconi_2017} have a cascade topology. None of them is track-like\footnote{The topology of IceCube detections can be 
broadly classified in two types: (1) cascade-like, characterised by a compact spherical 
energy deposition, which can only be reconstructed with a spatial resolution $\approx 15^{\circ}$; 
(2) track-like, defined by a dominant linear topology from the induced muon, 
with positions known typically within one degree or less.}. 
Nevertheless, 
after a long enough exposure a track IceCube signal from blazars,  if real, should also start to appear. This is of great interest because false (random) associations of 
tracks with a blazar are less likely due to the better defined position of this 
event-class with respect to cascades.

\cite{Brown_2015} and \cite{Palladino_2017} did address this issue by using tracks, with null results.
The former paper looked for $\gamma$-ray emission spatially coincident with the 37 IceCube tracks 
published by \cite{Aartsen2014} using 70 months of {\it Fermi}-Large Area Telescope (LAT) observations. 
The latter cross-correlated the 2FHL catalogue with the 29 IceCube tracks 
published by \cite{Aartsen2016}. 

Various recent results warrant a reappraisal of these topics. Namely: 1. the availability of many more 
IceCube tracks based on an updated version of the list given in \cite{outflows}. In this paper we use 70 tracks with positions off the Galactic plane (|b|$>$10$^{\circ}$) and angular uncertainty $\le 3^{\circ}$; 
2. the development of a new tool, "VOU-Blazar" 
\citep{dissecting,vou-blazar}, within the Open Universe initiative 
\citep{openuniverse} to find blazars and blazar candidates in relatively large areas of the sky 
on the basis of {\it all} the multi-frequency data available; 3. the release of newly processed multi-frequency 
data (especially for \swift: e.g., \citealt{Giommi_2019a}); 4. the availability of new catalogues of high-energy 
emitting blazars, e.g., the third high-synchrotron peaked (3HSP) \citep{3hsp}, 4FGL \citep{4FGL} and 
4LAC \citep{4LAC}; 5. the fact that TXS\,0506+056, with a typical \nup~ of $\approx 
10^{14.5}$~Hz, is not an HBL but rather an IBL that during flares reaches the HBL threshold \citep{dissecting}; 6. finally since, despite appearances, 
TXS\,0506+056 is {\it not} a blazar of the BL Lac type but instead a masquerading BL Lac, i.e., 
intrinsically a FSRQ with hidden broad lines \citep{Padovani_2019}, this also suggests that HBLs cannot be the full story.  
It is therefore now time to re-assess the issue of the possible match between blazars and 
IceCube tracks in a statistical fashion. We tackle this in two complementary ways: 1. a cross-matching of 
the IceCube events with catalogues of high-energy sources and of known blazars; 2. a detailed dissection
of each neutrino error region using the VOU-Blazars tool, which takes advantage of all the available
multi-frequency data. 

\section{The sample of IceCube neutrino events}\label{sec:sample}

\begin{figure}
\centering
\includegraphics[width=8.5cm]{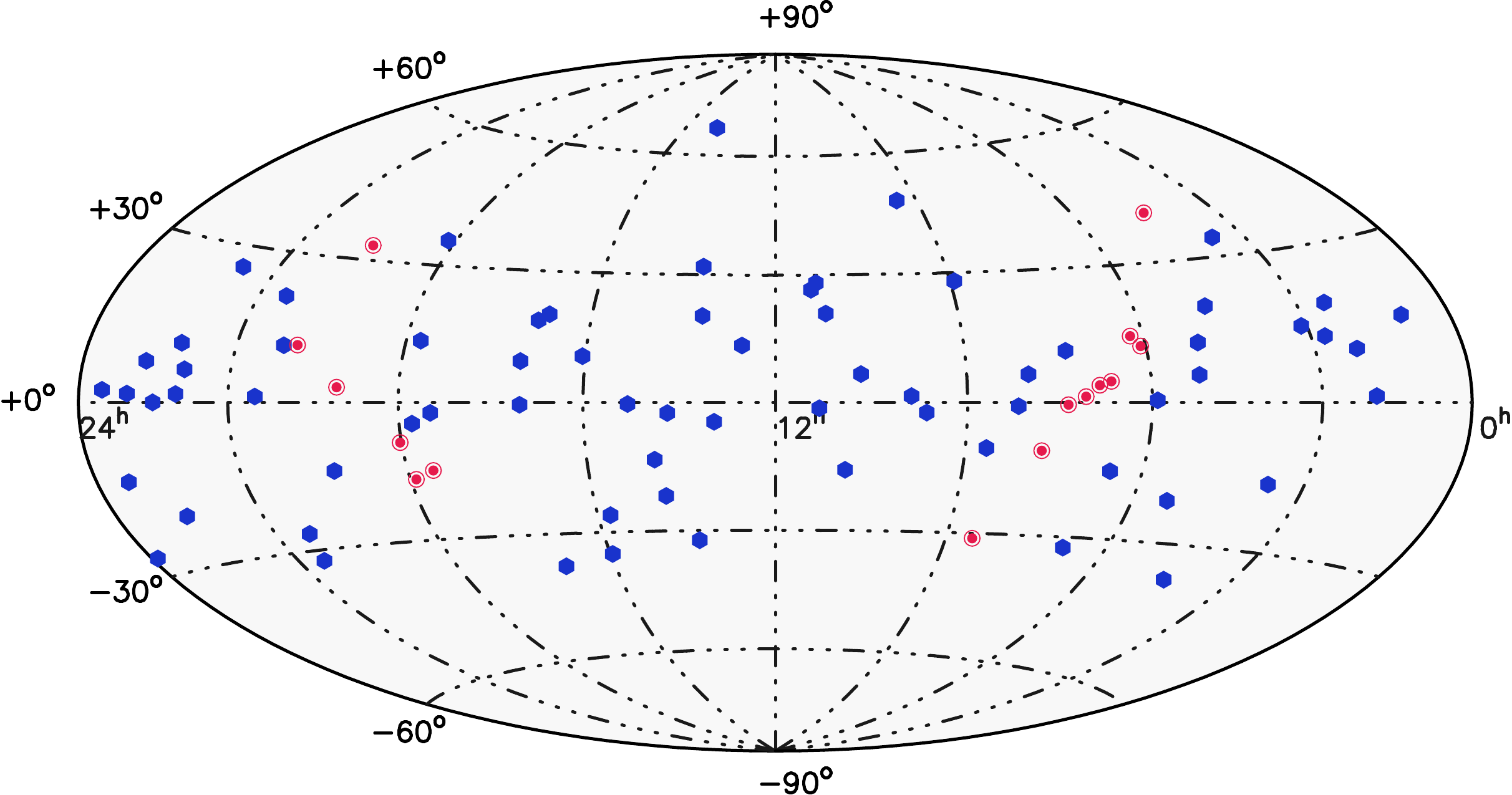}
\caption{The sky distribution of the sample of IceCube tracks plotted in equatorial coordinates and Hammer-Aitoff projection. Low Galactic latitude events (|b|$<$10$^{\circ}
$) are plotted as red symbols. The blue solid filled symbols represent the 70 higher latitudes tracks that have been considered for our statistics. Note that almost all tracks 
have declinations $|\delta| \leq 35^{\circ}$.}
\label{fig:aitoff}
\end{figure}

IceCube detects neutrinos in an energy range from $10\,\text{GeV}$ to several PeV \citep{Aartsen2016}. While most of the events are associated with the atmospheric background, some high-energy events have a good chance of being of astrophysical origin. The two most important channels for neutrino astronomy are starting- and through-going tracks. They mainly correspond to muon neutrinos doing charged-current interactions inside and outside the detector volume, respectively. Due to the longer lever arm compared to spherical cascades, tracks can reach an angular resolution as low as 1$^{\circ}$ \citep{Aartsen2017}. There is, however, some level of uncertainty about the detector systematic uncertainties and subsequently the reconstruction errors, which we compensate for by scaling the major and minor axes of the 90\% error ellipses,
$\Omega_{90}$, to 1.1, 1.3 and 1.5 times their original size 
($\Omega_{90\times1.1}, \Omega_{90\times1.3},  
 \Omega_{90\times1.5}$ respectively). 
  Note that IceCube applies different treatments to the angular and energetic systematic uncertainties: while some very high-energy events have been re-simulated including a complete ensemble of systematic errors  \citep{Aartsen2017,icfermi}, other archival events use a scaling of the test-statistic distribution  \citep{icfermi} or do not include systematic uncertainties at all \citep{Aartsen2017}. 
 Given that the quoted errors on the angular radii are different for RA and Dec, we approximate the region as an ellipse. Moreover, since the errors are also asymmetric, the centre of the ellipse was shifted in the direction of the most significant error by an amount equal to half the difference between the larger and smaller errors, and by setting the major and minor axes equal to the sum of the two asymmetrical errors.
 In the southern hemisphere IceCube is strongly dominated by atmospheric muons, hence the selection of neutrino-induced through-going tracks is limited to the northern hemisphere ($\delta >5^{\circ}$). In contrast, starting tracks provide an all-sky channel to search for astrophysical neutrino events. Despite those background considerations, the absorption of  high-energy neutrinos by the Earth effectively shrinks the field of view (FoV) to declinations approximately in the range $ -35^{\circ}$ to $ +35^{\circ}$, as shown in Figure \ref{fig:aitoff}. In this work we combine the list of highest-energy through-going tracks from IceCube's diffuse astrophysical muon-neutrino search (DIF), with the selection of high-energy starting tracks (HES) and the events published as alerts in the scope of IceCube's realtime program (AHES, EHE)  \citep{2013PhRvL.111b1103A,Aartsen2015,Aartsen2016,ICECube13,ICECube14,ICECube15_1,ICECube17_1,ICECube17_2,2019ICRC...36.1004S,2019ICRC...36.1017S}. The abbreviations in parenthesis refer to previous naming of the events and are given in Tables\,\ref{sample} and \ref{trackstable_G} for reference. After cutting out events with a poor angular resolution ( sources for which the area of the ellipse was larger than that of a circle with r $= 3^{\circ}$) or with the flag \textit{'bad angular resolution'} in \cite{datarelease}\footnote{The complete list is available under \url{https://icecube.wisc.edu/science/data/TXS0506_alerts}} and removing events close to the Galactic plane (|b|$<$10$^{\circ}$) the final sample contains 70 events. The angular resolution cut is motivated by the fact that just by random coincidence we expect to see one IBL or HBL \textit{and} one LBL counterpart candidate every $\sim$27 square degrees \citep{4LAC}, which is equivalent to a circle of radius 3$^{\circ}$. With the cut on the Galactic latitude we remove all the events for which it is hard to identify extra-galactic counterparts due to high source density, \gr\, source confusion, and foreground Galactic emission. Our final list is an updated version of the one used by \cite{outflows}. 
The complete sample of high-energy IceCube neutrino tracks is presented in Tables\,\ref{sample} (used events) and \ref{trackstable_G} (discarded events) where the respective neutrino names are given in columns 1 and 2, the Modified Julian Date (MJD) of arrival times is given in column 3, the positions in columns 4 and 5, and the Galactic latitude in column 6.

For a small fraction of events from the high-energy starting event sample (HES) only a fixed median angular resolution of $\sim 1.3^{\circ}$ is public. For the statistical analysis we treat them in the same manner as the other (90 per cent) error ellipses ($\Omega_{90}$), but do not write the error explicitly in Tables\,\ref{sample} and \,\ref{trackstable_G}.

\begin{table*}
\begin{center}
\caption{The sample of 70 IceCube tracks considered in this paper. Columns 1 and 2 give the standard IceCube name and previous namings, respectively. The other columns give the MJD, right ascension and declination with 90\% error (if available) and the Galactic latitude. Wherever no 90\% error is given we use a fixed \textit{median} angular resolution of 1.3 degrees \protect\citep{ICECube13, ICECube15_1, ICECube17_2}. The events are sorted by right ascension. Whenever \protect\cite{datarelease} provided an updated reconstruction we add a $\dagger$ to the event name.} 
\resizebox{\textwidth}{!}{
\def\arraystretch{1.3}
\begin{tabular}{L{3.cm}L{2.5cm}L{2.5cm}L{2.5cm}L{2.5cm}L{1.2cm}}
\hline
IceCube Name & Other IceCube Name & MJD & RA& Dec & Galactic     \\
& &  & J2000.0  & J2000.0  & Latitude \\ 
&  & & (deg) & (deg) & (deg)    \\ \hline \hline
IceCube-160331A & DIF35& 57478.60  & $15.60\,^{+0.45}_{-0.58}$ & $15.60\,^{+0.53}_{-0.60}$  & -47.19  \\ \hline
IceCube-090813A & DIF1& 55056.70  & $29.51\,^{+0.40}_{-0.38}$ & $1.23\,^{+0.18}_{-0.22}$  & -57.42  \\ \hline
IceCube-131014A & DIF23& 56579.91  & $32.94\,^{+0.63}_{-0.62}$ & $10.20\,^{+0.34}_{-0.49}$  & -47.90 \\ \hline
IceCube-111216A & DIF16& 55911.28  & $36.65\,^{+1.85}_{-1.71}$ & $19.10\,^{+2.21}_{-2.21}$  & -38.34  \\ \hline
IceCube-161103A & AHES4&  57695.38  &   $ 40.83\,^{+1.10}_{-0.70}$  & $12.56\,^{+1.10}_{-0.65}$  & -41.92 \\ \hline
IceCube-161210A & EHE3& 57732.84  & $46.58\,^{+1.10}_{-1.00}$ & $14.98\,^{+0.45}_{-0.40}$  & -36.67 \\ \hline
IceCube-150831A &  & 57265.22  & $54.85\,^{+0.94}_{-0.98}$ & $33.96\,^{+1.07}_{-1.19}$  & -17.09 \\ \hline
IceCube-141109A & HES61$\dagger$ & 56970.21  & $55.63\,^{+0.79}_{-1.53}$ & $-16.50\,^{+0.81}_{-0.68}$  & -49.11  \\ \hline
IceCube-190504A & & 58607.77  & $65.79\,^{+1.23}_{-1.23}$ & $-37.44\,^{+1.23}_{-1.23}$  & -44.68 \\  \hline
IceCube-120922A &  & 56192.55  & $70.75\,^{+1.56}_{-1.63}$ & $19.79\,^{+1.37}_{-0.68}$  & -16.90 \\ \hline
IceCube-151114A & DIF34& 57340.90  & $76.30\,^{+0.75}_{-0.74}$ & $12.60\,^{+0.61}_{-0.58}$  & -16.79  \\ \hline
IceCube-170922A & EHE5 & 58018.87  & $77.43\,^{+0.95}_{-0.65}$ & $5.72\,^{+0.50}_{-0.30}$  & -19.56 \\ \hline
IceCube-150428A & HES71$\dagger$ & 57140.47  & $80.77\,^{+1.12}_{-1.23}$ & $-20.75\,^{+0.45}_{-0.83}$  & -28.33  \\ \hline
IceCube-101028A & DIF9$\dagger$ & 55497.30  & $88.68\,^{+0.54}_{-0.55}$ & $0.46\,^{+0.33}_{-0.27}$  & -12.38  \\ \hline
IceCube-170321A & EHE4 & 57833.31  & $98.30\,^{+1.20}_{-1.20}$ & $-15.02\,^{+1.20}_{-1.20}$  & -10.75  \\ \hline
IceCube-140721A & HES58& 56859.76  & $102.10$ & $-32.40$  & -14.73 \\  \hline
IceCube-140611A & DIF27$\dagger$ & 56819.20  & $110.30\,^{+0.66}_{-0.45}$ & $11.57\,^{+0.14}_{-0.24}$  & 11.79 \\ \hline
IceCube-190503A& & 58606.72  & $120.28\,^{+0.57}_{-0.77}$ & $6.35\,^{+0.76}_{-0.70}$  & 18.37  \\  \hline
IceCube-160806A & EHE2 & 57606.51  & $122.81\,^{+0.50}_{-0.50}$ & $-0.81\,^{+0.50}_{-0.50}$  & 17.29 \\ \hline
IceCube-130907A & & 56542.79  & $129.81\,^{+0.48}_{-0.28}$ & $-10.36\,^{+0.36}_{-0.31}$  & 18.35   \\   \hline
IceCube-150904A & DIF32& 57269.80  & $134.00\,^{+0.39}_{-0.58}$ & $28.00\,^{+0.47}_{-0.47}$  & 38.35 \\ \hline
IceCube-100623A & DIF4& 55370.74  & $141.25\,^{+0.46}_{-0.45}$ & $47.80\,^{+0.56}_{-0.48}$  & 45.16   \\ \hline
IceCube-180908A &  & 58369.83  & $144.58\,^{+1.55}_{-1.45}$ & $ -2.13\,^{+0.9}_{-1.2}$  & 35.09 \\  \hline
IceCube-141209A & HES63$\dagger$& 57000.14  & $160.05\,^{+0.84}_{-1.04}$ & $6.57\,^{+0.64}_{-0.56}$  & 52.68     \\ \hline
IceCube-171015A &  & 58041.07  & $162.86\,^{+2.60}_{-1.70}$ & $ -15.44\,^{+1.60}_{-2.00}$  & 38.43    \\ \hline
IceCube-130408A & HES37 & 56390.19  & $167.17\,^{+2.87}_{-1.90}$ & $20.67\,^{+1.15}_{-0.89}$  & 65.69   \\ \hline
IceCube-121026A & DIF20& 56226.60  & $169.61\,^{+1.16}_{-1.11}$ & $28.04\,^{+0.67}_{-0.66}$  & 69.40\\ \hline
IceCube-140923A & & 56923.72  & $169.72\,^{+0.91}_{-0.86}$ & $-1.34\,^{+0.73}_{-0.66}$  & 53.85  \\  \hline
IceCube-120523A & & 56070.57  & $171.03\,^{+0.81}_{-0.90}$ & $26.36\,^{+0.49}_{-0.30}$  & 70.51   \\   \hline
IceCube-190819A & & 56070.57  & $148.80\,^{+2.07}_{-3.24}$ & $1.38\,^{+1.00}_{-0.75}$  & 70.51  \\ \hline
IceCube-141126A & HES62& 56987.77  & $187.90$ & $13.30$  & 75.41 \\  \hline 
IceCube-150926A & & 57291.90  & $194.50\,^{+0.76}_{-1.21}$ & $-4.34\,^{+0.70}_{-0.95}$  & 58.49\\ \hline
IceCube-151017A & DIF33& 57312.70  & $197.60\,^{+2.46}_{-2.09}$ & $19.90\,^{+2.82}_{-2.21}$  & 81.57\\ \hline
IceCube-120515A & DIF17& 56062.96  & $198.74\,^{+1.44}_{-1.09}$ & $31.96\,^{+0.81}_{-0.85}$  & 82.97\\ \hline
IceCube-160814A & AHES3& 57614.91  & $200.30\,^{+2.43}_{-3.03}$ & $-32.40\,^{+1.39}_{-1.21}$  & 30.05\\ \hline
IceCube-121011A & DIF19$\dagger$ & 56211.77  & $205.22\,^{+0.59}_{-0.65}$ & $-2.39\,^{+0.51}_{-0.57}$  & 58.17  \\ \hline
IceCube-131202A & HES43$\dagger$ & 56628.57  & $206.63\,^{+2.04}_{-1.56}$ & $-22.02\,^{+1.69}_{-1.04}$  & 39.08 \\ \hline
IceCube-120123A & HES23& 55949.57  & $208.70$ & $-13.20$  & 46.84 \\  \hline
IceCube-140216A & HES47& 56704.60  & $209.40$ & $67.40$  & 48.49 \\ \hline
\end{tabular}}
\label{sample}
\end{center}
\end{table*}

\begin{table*}
\ContinuedFloat
\begin{center}
\caption{continued} 
\resizebox{\textwidth}{!}{
\def\arraystretch{1.3}
\begin{tabular}{L{3.cm}L{2.5cm}L{2.5cm}L{2.5cm}L{2.5cm}L{1.2cm}}
\hline
IceCube Name & Other IceCube Name & MJD & RA& Dec & Galactic     \\
& &  & J2000.0  & J2000.0  & Latitude \\ 
&  & & (deg) & (deg) & (deg)    \\  \hline \hline
IceCube-160731A & EHE1, AHES2 & 57600.08  & $214.50\,^{+0.75}_{-0.75}$ & $-0.33\,^{+0.75}_{-0.75}$  & 55.55 \\ \hline
IceCube-170506A & AHES6 & 57879.53  & $221.80\,^{+3.00}_{-3.00}$ & $-26.00\,^{+2.00}_{-2.00}$  & 30.01 \\ \hline
IceCube-130817A & DIF22 & 56521.83  & $224.89\,^{+0.87}_{-1.19}$ & $-4.44\,^{+1.21}_{-0.94}$  & 45.80  \\  \hline 
IceCube-181014A & & 58405.50 & $225.15\,^{+1.40}_{-2.85}$ & $-34.80\,^{+1.15}_{-1.85}$  & 20.95 \\ \hline
IceCube-190730A & & 58694.87  & $225.79\,^{+1.28}_{-1.43}$ & $10.47\,^{+1.14}_{-0.89}$  & 54.83 \\  \hline
IceCube-110521A & DIF12& 55702.77  & $235.13\,^{+2.70}_{-1.76}$ & $20.30\,^{+1.00}_{-1.43}$  & 50.88 \\ \hline
IceCube-120301A & & 55987.81  & $238.01\,^{+0.60}_{-0.59}$ & $18.60\,^{+0.46}_{-0.39}$  & 47.76  \\  \hline 
IceCube-140420A & HES53$\dagger$& 56767.07  & $238.98\,^{+1.81}_{-1.91}$ & $-37.73\,^{+1.47}_{-1.31}$  & 12.06 \\ \hline
IceCube-150911A & HES76$\dagger$ & 57276.57  & $240.20\,^{+1.29}_{-1.38}$ & $-0.45\,^{+1.17}_{-1.23}$  & 36.83\\  \hline
IceCube-160427A & AHES1, HES82$\dagger$& 57505.25  & $240.57\,^{+0.60}_{-0.60}$ & $9.34\,^{+0.60}_{-0.60}$  & 41.68 \\ \hline
IceCube-151122A & & 57348.53  & $262.18\,^{+0.90}_{-1.21}$ & $-2.38\,^{+0.73}_{-0.43}$  & 17.16  \\ \hline
IceCube-110930A & & 55834.45  & $266.48\,^{+2.09}_{-1.55}$ & $-4.41\,^{+0.59}_{-0.86}$  & 12.43  \\ \hline
IceCube-100925A & DIF7& 55464.90  & $266.29\,^{+0.58}_{-0.62}$ & $13.40\,^{+0.52}_{-0.45}$  & 20.64\\ \hline
IceCube-110610A & DIF13& 55722.43  & $272.22\,^{+1.23}_{-1.19}$ & $35.55\,^{+0.69}_{-0.69}$  & 23.50 \\ \hline
IceCube-131204A & & 56630.47  & $289.16\,^{+1.08}_{-0.94}$ & $-14.25\,^{+0.91}_{-0.81}$  & -11.94 \\ \hline 
IceCube-131023A & & 56588.56  & $301.82\,^{+1.10}_{-0.93}$ & $11.49\,^{+1.19}_{-1.09}$  & -11.10\\ \hline
IceCube-170312A & AHES5 & 57824.58  & $305.15\,^{+0.50}_{-0.50}$ & $-26.61\,^{+0.50}_{-0.50}$  & -30.40  \\ \hline
IceCube-100710A & DIF5& 55387.54  & $306.96\,^{+2.70}_{-2.28}$ & $21.00\,^{+2.25}_{-1.56}$  & -10.13\\ \hline
IceCube-190124A & & 58507.15  & $307.40\,^{+0.80}_{-0.90}$ & $-32.18\,^{+0.70}_{-0.70}$  & -33.76 \\ \hline
IceCube-110128A & DIF11$\dagger$ & 55589.56  & $307.53\,^{+0.82}_{-0.81}$ & $1.19\,^{+0.35}_{-0.32}$  & -21.22 \\ \hline
IceCube-150714A & DIF30& 57217.90  & $325.50\,^{+1.77}_{-1.46}$ & $26.10\,^{+1.68}_{-1.85}$  & -19.93 \\ \hline
IceCube-150812A & DIF31$\dagger$ & 57246.76  & $328.19\,^{+1.01}_{-1.03}$ & $6.21\,^{+0.44}_{-0.49}$  & -35.44\\  \hline
IceCube-120807A & DIF18& 56146.21  & $330.10\,^{+0.65}_{-0.82}$ & $1.57\,^{+0.46}_{-0.42}$  & -39.84  \\ \hline
IceCube-101009A & DIF8$\dagger$ & 55478.38  & $331.09\,^{+0.56}_{-0.72}$ & $11.10\,^{+0.48}_{-0.58}$  & -34.30  \\   \hline
IceCube-140114A & HES44& 56671.88  & $336.71$ & $0.04$  & -45.92 \\  \hline  
IceCube-190331A &  & 58573.29  & $337.68\,^{+0.23}_{-0.34}$ & $-20.70\,^{+0.30}_{-0.48}$  & -57.31\\ \hline
IceCube-171106A &  & 58063.77  & $340.00\,^{+0.7}_{-0.5}$ & $ 7.40\,^{+0.35}_{-0.25}$  & -43.05 \\  \hline
IceCube-140108A &  & 56665.31  & $344.53\,^{+0.67}_{-0.48}$ & $1.57\,^{+0.35}_{-0.32}$  & -50.41 \\  \hline
IceCube-140203A & & 56691.79  & $349.54\,^{+2.21}_{-1.97}$ & $-13.71\,^{1.23}_{-1.38}$  & -64.43  \\ \hline
IceCube-160510A &  & 57518.66  & $352.34\,^{+1.63}_{-1.31}$ & $2.09\,^{+0.99}_{-0.85}$  & -54.72\\  \hline
IceCube-190104A  &  & 58487.36 & $357.98 \,^{+2.30}_{-2.10}$ & $-26.65\,^{+2.20}_{-2.50}$  & -76.73 \\  \hline
\end{tabular}}
\end{center}
\end{table*}

\begin{table*}
\begin{center}
\caption{Additional list of 24 IceCube tracks that are close to the Galactic plane ($|b_{\rm II}| \le 10^{\circ}$) or {\bf for which the area of the error ellipse was larger than that of a circle with r $= 3^{\circ}$}. For these tracks no counterparts are given as the search for extra-galactic multi-wavelength sources in the background of the Galactic plane is {\bf not straightforward}. Whenever \protect\cite{datarelease} provided an updated reconstruction we add a $\dagger$ to the event name.} 
\resizebox{\textwidth}{!}{
\def\arraystretch{1.3}
\begin{tabular}{L{3.cm}L{2.5cm}L{2.5cm}L{2.5cm}L{2.5cm}L{1.2cm}}
\hline
IceCube Name & Other IceCube Name & MJD & RA& Dec & Galactic     \\
& &  & J2000.0  & J2000.0  & Latitude \\ 
&  & & (deg) & (deg) & (deg)    \\ \hline \hline
IceCube-110714A & HES13$\dagger$ & 55756.11  & $67.86\,^{+0.51}_{-0.72}$ & $40.32\,^{+0.73}_{-0.25}$  & -5.40  \\ \hline 
IceCube-190712A &  & 58676.05  & $76.46\,^{+5.09}_{-6.83}$ & $13.06 \,^{+4.48}_{-3.44}$  & -16.40  \\  \hline
IceCube-150515A & DIF29& 57157.94  & $91.60\,^{+0.16}_{-0.74}$ & $12.18\,^{+0.37}_{-0.35}$  & -4.22 \\ \hline
IceCube-130627A & DIF21, HES38 & 56470.11  & $93.38\,^{+0.83}_{-0.90}$ & $14.46\,^{+0.86}_{-0.94}$  & -1.61 \\ \hline
IceCube-150127A & DIF28& 57049.48  & $100.48\,^{+0.95}_{-1.87}$ & $4.56\,^{+0.68}_{-0.50}$  & -0.03   \\ \hline 
IceCube-150923A &  & 57288.03  & $103.27\,^{+0.70}_{-1.36}$ & $3.88\,^{+0.59}_{-0.71}$  & 2.13 \\ \hline
IceCube-140609A  & DIF26& 56817.64  & $106.26\,^{+2.27}_{-1.90}$ & $1.29\,^{+0.83}_{-0.74}$  & 3.61  \\ \hline
IceCube-101112A & HES5& 55512.55  & $110.56\,^{+0.80}_{-0.37}$ & $-0.37\,^{+0.48}_{-0.65}$  & 6.68 \\\hline
IceCube-110304A & & 55624.95  & $116.37\,^{+0.73}_{-0.73}$ & $-10.72\,^{+0.57}_{-0.65}$  & 6.86  \\ \hline
IceCube-100912A & HES3 & 55451.07  & $127.90$ & $-31.20$  & 4.93 \\  \hline
IceCube-190704A & & 58668.78  & $161.85\,^{+2.16}_{-4.33}$ & $27.11\,^{+1.81}_{-1.83}$  & 62.47  \\   \hline
IceCube-140122A & HES45 & 56679.21  & $219.64\,^{+5.16}_{-4.16}$ & $-86.16\,^{+0.55}_{-0.60}$  & -23.69  \\ \hline 
IceCube-111201A & DIF15& 55896.86  & $222.87\,^{+1.95}_{-7.73}$ & $1.87\,^{+1.25}_{-1.18}$  & 51.73  \\ \hline
IceCube-100813A & DIF6 & 55421.51  & $252.00\,^{9.56}_{-16.65}$ & $15.21\,^{9.35}_{-7.41}$  & 34.07  \\ \hline
IceCube-160128A & & 57415.18  & $263.40\,^{+1.35}_{-1.18}$ & $-14.79\,^{+0.99}_{-1.02}$  & 9.80  \\ \hline
IceCube-190221A &  & 58535.35  & $268.81\,^{+1.20}_{-1.80}$ & $-17.04\,^{+1.30}_{-0.5}$  & 4.18 \\ \hline
IceCube-181023A &  & 58414.69 & $270.18\,^{+2.00}_{-1.70}$ & $ -8.57 \,^{+1.25}_{-1.30}$  & 7.19 \\ \hline
IceCube-101113A& DIF10& 55513.60  & $285.95\,^{1.29}_{-1.50}$ & $3.15\,^{+0.70}_{-0.63}$  & -1.31 \\ \hline 
IceCube-140109A & DIF24$\dagger$ & 56666.50  & $292.85\,^{+0.87}_{-0.94}$ & $33.06\,^{+0.50}_{-0.46}$  & 6.85\\ \hline
IceCube-091106A & DIF2& 55141.13  & $298.21\,^{+0.53}_{-0.57}$ & $11.74\,^{+0.32}_{-0.38}$  & -7.93 \\ \hline
IceCube-110722A & DIF14& 55764.22  & $315.66\,^{+5.91}_{-5.35}$ & $5.29\,^{+4.85}_{-4.72}$  & -25.98   \\ \hline
IceCube-190619A &  & 58653.65  & $343.26\,^{+ 4.08}_{- 2.63}$ & $10.73\,^{+1.51}_{-2.61}$  & -57.69 \\  \hline
IceCube-100608A & DIF3 & 55355.49  & $344.93\,^{+3.39}_{-2.90}$ & $23.58\,^{+2.31}_{-4.13}$  & -32.57  \\ \hline
IceCube-140522A & DIF25& 56799.96  & $349.39\,^{2.89}_{-4.12}$ & $18.05\,^{1.94}_{-1.80}$  & -39.41\\  \hline
\end{tabular}}
\label{trackstable_G}
\end{center}
\end{table*}

\section{Searching for \gr\, blazars in IceCube neutrino error regions}

\subsection{Cross-matching with catalogues of \gr\,sources}\label{sec:statan}
Since the production of neutrinos is accompanied by high-energy $\gamma$-rays
the obvious choice for looking for possible matches with IceCube tracks is given 
by $\gamma$-ray catalogues, or specific samples of blazars that are expected to emit in the \gr\, band.
We then used the following catalogues: the {\it Fermi}-LAT 3FHL \citep{3FHL}, 4LAC \citep{4LAC}, and the multi-wavelength based 3HSP \citep{3hsp}. 
While the first two are catalogues of \gr\,sources, about half of the blazars in the 3HSP catalogue have not been detected yet by {\it Fermi}-LAT as individual sources. However, \cite{Paliya} have found a very strong signal ($> 32\,\sigma$ confidence) in the {\it Fermi}-LAT stacking analysis of the sample of  still \gr\,undetected 
3HSP sources with \nup~$>10^{17}$~Hz.

Based on these  catalogues, we perform a statistical analysis similar to the one originally presented in \cite{Padovani_2016}, who found a first hint of correlation between IceCube neutrinos and extreme blazars at a significance level of $\sim 1.4$ per cent. 

We estimate the total number of sources lying inside $\Omega_{90}$ ($N_{s}$) and compare it to the expectation from randomised samples. The approximately elliptical shape of the contours is preserved in the test.
For each of the catalogues, the chance probability to observe a certain $N_{s}$ is calculated using a set of $10^4$ randomised realisations of the catalogues.
To preserve the distribution of the extra-galactic sources and have a representative set of random cases, a random Galactic longitude value is assigned to each source.
When available we also follow the classification of sources provided within the catalogues. We report in Table\,\ref{statantable} the results of the tests performed. We applied a correction for the ``look elsewhere effect'' since we perform multiple tests.
The corresponding p-value and its significance in units of $\sigma$ of the normal distribution is reported once for each hypothesis, and once for the complete set of tests.
 
The results of the statistical tests can be summarised as follows:
\begin{itemize}
    \item the smallest p-value of $10^{-3}$ is obtained for the 3HSP catalogue with 29 sources over an expected background of $\sim$\,16 sources. After trial-correction this corresponds to a significance of 2.79 $\sigma$ in a one-sided normal distribution;
    \item no other excess is observed in the other catalogues and classes of objects.
\end{itemize}
In Table\,\ref{corr_nogamma_table}, we report the list of 3HSP objects within $\Omega_{90}$ (90 per cent error region) without a \gr\, counterpart. All the objects \textit{with} \gr\, detections, also found using the VOU-Blazars tool, are  discussed in Section \ref{sec:vou} and are listed in Table\,\ref{trackstable}. 

We note that: 1. the 3HSP catalogue includes, by 
definition, only HBL and excludes, for example, TXS\,0506+056-like blazars; 2. the 
3FHL catalogue has a cut-off at 10\,GeV, which implies a reduced sensitivity because of 
the narrower LAT band; moreover, although we have re-derived all \nup~values, $\sim 
7$ per cent of the sources are still missing it; 3. only $\sim$\,76 per cent of the 
4LAC sources have a value of \nup, which is in any case derived using the 
(still limited) number of catalogues available through the Space Science Data Center 
(SSDC) SED builder tool\footnote{\url{https://tools.ssdc.asi.it/SED/}}
and a method to disentangle the jet from other SED components (host galaxy, blue 
bump, etc.), which is different from the one we use, as detailed below (Sect. \ref{sec:vou}). 
To overcome these intrinsic limitations of existing catalogues, we proceed with a complete dissection of the regions around IceCube neutrinos and of the objects 
therein contained.

\begin{table*}
\begin{center}
\caption{Results of the statistical analysis described in details in Section \ref{sec:statan}. The table presents the measured and expected number
of counterparts in $\Omega_{90}$ and the relative p-values.}
\begin{tabular}{llcccc}
\toprule
Catalog &Source Class & Number of Sources & Expected $N_s$ & Measured $N_s$ & p-value\\
\midrule
\midrule
\textbf{4LAC} &&&&\\
\midrule
& All Blazars & 2794 & 21.6 & 26 & 0.20\\
& LSP & 1289 & 9.9 & 10 & 0.53\\
& ISP+HSP & 843 & 6.8 & 8 & 0.37\\
& No SED Class. & 662 & 5.0 & 8 & 0.13\\
\midrule
\multicolumn{6}{c}{Post trial p-value: 0.38 (0.30$\sigma$)}\\
\midrule
\textbf{3FHL} &&&&\\
\midrule
& All Blazars & 1301 & 9.3 & 14 & 0.09\\
& LSP & 400 & 3.1 & 5 & 0.19\\
& ISP+HSP & 901 & 6.3 & 9 & 0.18\\
\midrule
\multicolumn{6}{c}{Post trial p-value: 0.15 (1.02$\sigma$)}\\
\midrule
\textbf{3HSP} &&&&\\
\midrule
& All & 2011 & 15.8 & 29 & 10$^{-3}$\\
& $\gamma$-ray detected & 1005 & 7.7 & 12 & 9$\times $10$^{-2}$\\
\midrule
\multicolumn{6}{c}{Post trial p-value: 3$\times $10$^{-3}$ (2.79$\sigma$)}\\
\midrule
\bottomrule
\end{tabular}
\label{statantable}
\end{center}
\end{table*}

\begin{table*}
\begin{center}
\caption{The 3HSP objects situated inside the
angular uncertainty $\Omega_{90}$ of the IceCube neutrino events (see Sect.
\ref{sec:statan}). Only objects not detected in $\gamma$-rays are listed, see Table \ref{trackstable} for the $\gamma$-ray detected ones}
\begin{tabular}{cccc}
\toprule
\textbf{IceCube Name} & \textbf{Object Name} & \textbf{IceCube Name} & \textbf{Object Name}\\
\midrule
\midrule
IceCube-140216A & 3HSP J140203.8+674104 & IceCube-101009A & 3HSP J220214.9+104130 \\
\cmidrule(lr){1-2} \cmidrule(lr){3-4}
IceCube-170506A & 3HSP J144437.0-250931 & IceCube-140203A & 3HSP J231752.1-144324 \\
\cmidrule(lr){3-4}
                & 3HSP J145021.4-273052 & IceCube-180908A & 3HSP J093938.5-031503 \\
\cmidrule(lr){1-2} \cmidrule(lr){3-4}
IceCube-111216A & 3HSP J023006.0+194921 & IceCube-190104A & 3HSP J235023.3-243602 \\
\cmidrule(lr){1-2} \cmidrule(lr){3-4}
IceCube-151017A & 3HSP J130631.0+192244 & IceCube-190730A & 3HSP J150604.4+102233 \\
\cmidrule(lr){3-4}
                & 3HSP J131639.8+205514 & IceCube-190819A & 3HSP J095127.8+010210 \\
\cmidrule(lr){1-2}
IceCube-110521A & 3HSP J154939.8+195355 &                 & 3HSPJ095649.5+015601 \\
\cmidrule(lr){1-2}
IceCube-100925A & 3HSP J174442.8+134802 &                 & 3HSPJ095849.0+013219 \\
\cmidrule(lr){1-2}
IceCube-170922A & 3HSP J050833.4+053109 \\
\midrule
\bottomrule
\end{tabular}
\label{corr_nogamma_table}
\end{center}
\end{table*}

\begin{table*}
\begin{center}
\caption{Table of IceCube tracks with possible counterparts within 1.5 times the 90\% error ellipses ($\Omega_{90\times1.5}$). In addition to the source names we also give the $\rm 1.4$\,GHz radio flux, the \nup\ for the SED classification and the redshift. Bold \textbf{event names} 
indicate tracks with at least one plausible counterpart, while the letters a, b, c, d indicate whether the source was found in $\Omega_{90}$, $\Omega_{90\times1.1}$, $\Omega_{90\times1.3}$, $\Omega_{90\times1.5}$ respectively.} 
\resizebox{\textwidth}{!}{
\def\arraystretch{1.3}
\begin{tabular}{L{4.0cm}L{5.0cm}L{2.cm}L{2.cm}L{2.cm}}
\hline
IceCube Name & Source Name &  $S_{\rm 1.4 GHz}$ & \nup & Redshift\\
 &  & [mJy] & [Hz]  &      \\ \hline \hline
\setrow{\bfseries} IceCube-160331A  & 3HSP J010326.0+152624 $^{\text{a}}$ &  225 & 15.0 &   0.246 \\ \hline
\setrow{\bfseries}IceCube-090813A &  5BZU J0158+0101 $^{\text{b}}$  & 82  & 14.1 & 0.4537 \\ \hline
IceCube-131014A & & & & \\ \hline
\setrow{\bfseries}IceCube-111216A & 5BZQ J0225+1846 $^{\text{a}}$    &  461  & 12.5 &  2.69 \\ 
&   3HSP J023248.5+201717 $^{\text{a}}$   & 82  & 18.5 &  0.139 \\
&   VOU J022411+161500 $^{\text{d}}$   &   13 & 14.5  & 0.3 \\ \hline
\setrow{\bfseries}IceCube-161103A & VOU J024445+132002 $^{\text{a}}$ & 200 & 14.5 & 0.90 \\ 
& 3HSP J023927.2+132738 $^{\text{d}}$ & 20 & 15.0 &0.5 \\ \hline
IceCube-161210A & & & & \\ \hline
\setrow{\bfseries} IceCube-150831A &  3HSP J034424.9+343017 $^{\text{c}}$  & 13 & 15.7 &  --- \\
 & 5BZQ J0336+3218 $^{\text{d}}$ & 2677 & 12. & 1.26  \\   \hline
\setrow{\bfseries} IceCube-141109A & 3HSP J033913.6-173600 $^{\text{d}}$ & 171 &  15.6 & 0.065559 \\   \hline
\setrow{\bfseries}IceCube-190504A &  5BZB J0428-3756 $^{\text{a}}$  & 753 & 12.8 & 1.11 \\   
& 4LAC J0420.3-3745 $^{\text{a}}$ & 60 & <13.5 & 0.3 \\   \hline
IceCube-120922A &   &    &     &   \\ \hline
\setrow{\bfseries}IceCube-151114A & 5BZB J0502+1338 $^{\text{d}}$  & 545 & 13.2 & --  \\   \hline
\setrow{\bfseries}IceCube-170922A & 5BZB J0509+0541 $^{\text{a}}$   & 536 & 14.5. & 0.3365 \\ \hline
\setrow{\bfseries}IceCube-150428A & VOU J052526-201054 $^{\text{c}}$  & 231 &14.5 & 0.12 \\ \hline
IceCube-101028A   &   &  &  &   \\ \hline
\setrow{\bfseries}IceCube-170321A & 3HSP J062753.3-151957 $^{\text{c}}$   & 43 & 17.3 & 0.3102 \\ \hline
\setrow{\bfseries}IceCube-140721A & 3HSP J064933.5-313920 $^{\text{a}}$   & 8 & 17.0 & >0.563 \\   
 & 5BZQ J0648-3044 $^{\text{c}}$ & 898 & 12.5 &  1.15 \\  \hline
IceCube-140611A   &   &                 &                &   \\ \hline
IceCube-190503A   &   &                 &                &   \\  \hline
IceCube-160806A   &   &                 &                &   \\ \hline
IceCube-130907A   &   &                 &                &   \\   \hline
\setrow{\bfseries}IceCube-150904A &  3HSP J085410.1+275421 $^{\text{a}}$ &  15 & 16.1 & 0.4937 \\ \hline
IceCube-100623A   &   &                 &                &   \\ \hline
IceCube-180908A   &   &                 &                &   \\  \hline
\setrow{\bfseries}IceCube-141209A & VOU J104031+061721 $^{\text{a}}$ & 35 & 14.5 & --  \\
 & 5BZB J1043+0653 $^{\text{b}}$ & 8 & 14.5  &  0.4 \\ \hline
\setrow{\bfseries}IceCube-171015A & VOU J105603-180929 $^{\text{d}}$ & 12 & 14.1 & --    \\ \hline
\setrow{\bfseries} IceCube-130408A & 3HSP J111706.2+201407 $^{\text{a}}$ & 103 & 16.5  & 0.138 \\
 & 5BZQ J1059+2057 $^{\text{b}}$ & 121  & 13.0 & 0.39 \\
 & 3HSP J112405.3+204553 $^{\text{d}}$ & 9 & 15.3 & 0.54 \\
 & 3HSP J112503.6+214300 $^{\text{d}}$ & 8 & 15.8 & 0.36\\ \hline
IceCube-121026A   &   &                 &                &   \\ \hline
IceCube-140923A   &   &                 &                &   \\  \hline
\setrow{\bfseries}IceCube-120523A & 5BZQ J1125+2610 $^{\text{c}}$   & 921 & 12.5 &  2.34 \\   \hline
\end{tabular}}
\label{trackstable}
\end{center}
\end{table*}

\begin{table*}
\ContinuedFloat
\begin{center}
\caption{continued} 
\resizebox{\textwidth}{!}{
\def\arraystretch{1.3}
\begin{tabular}{L{4.0cm}L{5.0cm}L{2.cm}L{2.cm}L{2.cm}}
\hline
IceCube Name & Source Name &  $S_{\rm 1.4 GHz}$ & \nup & Redshift\\
 &  & [mJy] & [Hz]  &      \\ \hline \hline
\setrow{\bfseries}IceCube-190819A & 3HSP J094620.2+010451 $^{\text{a}}$   & 15 & $>18$ & 0.5768 \\
& 3HSP J100326.6+020455 $^{\text{c}}$   & 6 & 15.8 & 0.48 \\ 
& 5BZQ J0948+0022 $^{\text{d}}$ & 70 & 12.8 &  0.59 \\ \hline
\setrow{\bfseries}IceCube-141126A &  M87 $^{\text{a}}$  & 138488  & -- &  0.004\\   
& 3HSP J123123.9+142124 $^{\text{a}}$  & 54 & 16.0 &   0.25580 \\    \hline 
\setrow{\bfseries}IceCube-150926A & 3HSP J125848.0-044745 $^{\text{a}}$   & 4 & 17.0 &  0.586 \\   \hline
\setrow{\bfseries}IceCube-151017A & 5BZB J1314+2348 $^{\text{d}}$    &  184 & $\geq$14 & 0.15? \\
& 3HSP J130008.5+175537 $^{\text{d}}$  &  16 & 14.5 &  0.55 \\
& 5BZQ J1321+2216 $^{\text{d}}$  &  314 & 12.0 & 0.943 \\ 
& 3HSP J125821.5+212351 $^{\text{d}}$ & 26 & 16.7 & 0.6265\\ \hline
\setrow{\bfseries}IceCube-120515A  & 5BZU J1310+3220 $^{\text{b}}$  &  1687 & 12.5 & 0.997  \\
 & 5BZQ J1310+3233 $^{\text{b}}$   &  374 & 12.0 & 1.64 \\
 & 5BZB J1322+3216 $^{\text{c}}$    &  906 & 14.5 &  --\\ \hline
\setrow{\bfseries} IceCube-160814A &  5BZQ J1316-3338 $^{\text{b}}$  &  1277  & 12.5 & 1.21 \\ \hline
\setrow{\bfseries}IceCube-121011A & 5BZQ J1340-0137 $^{\text{d}}$ & 175 & 13.0 & 1.62   \\ \hline
\setrow{\bfseries} IceCube-131202A & 5BZQ J1342-2051 $^{\text{a}}$ &  399  & 12.0 & 1.58 \\ \hline
\setrow{\bfseries}IceCube-120123A &  VOU J135921-115043 $^{\text{d}}$  & 48 & 14.0 & 0.27 \\ \hline
\setrow{\bfseries}IceCube-140216A & 3HSP J140449.6+655431 $^{\text{c}}$    & 15 & 16.0 & 0.3627 \\    
& 5BZQ J1344+6606 $^{\text{d}}$  & 639 & 12.3 &  1.35 \\ \hline
IceCube-160731A & & & & \\ \hline
\setrow{\bfseries}IceCube-170506A &   3HSP J144656.8-265658 $^{\text{a}}$  &  41  & 17.6 & 0.32  \\
 &  VOU J143934-252458 $^{\text{a}}$ &   35 &  14.0 & 0.18 \\
 & 3HSP J143959.4-234140 $^{\text{c}}$  &  101  & 16.2 & 0.25  \\  \hline
IceCube-130817A & & & &  \\  \hline 
\setrow{\bfseries}IceCube-181014A  & 5BZB J1505-3432 $^{\text{a}}$  &  138 & 12.5 & --  \\ 
& 5BZQ J1457-3539 $^{\text{a}}$  &  675  & 13.5 & 1.42 \\ 
& VOU J150720-370902 $^{\text{d}}$ &  74  & 14.5 & -- \\ \hline
\setrow{\bfseries}IceCube-190730A  & 5BZQ J1504+1029 $^{\text{a}}$  &  1775 & 12.8 & 1.84  \\ \hline
\setrow{\bfseries}IceCube-110521A & 3HSP J155424.1+201125 $^{\text{c}}$  & 80 & 17.3  & 0.22227 \\
 & 3HSP J153311.2+185429 $^{\text{d}}$   & 23 & 17.0 & 0.305 \\
& 3HSP J152835.7+200420 $^{\text{d}}$  & 5 & 16.2 &  0.52 \\ \hline
IceCube-120301A & & & & \\  \hline 
IceCube-140420A & & &  & \\ \hline
\setrow{\bfseries} IceCube-150911A &  5BZQ J1557-0001 $^{\text{a}}$  &  1107  & 12.2 & 1.77\\  \hline
IceCube-160427A & & & & \\ \hline
IceCube-151122A & & & &  \\ \hline
\setrow{\bfseries}IceCube-110930A & 5BZQ J1743-0350 $^{\text{b}}$ & 1411 & 12.5 & 1.06 \\ \hline
IceCube-100925A & & & &\\ \hline
\setrow{\bfseries}IceCube-110610A & VOU J180812+350104 $^{\text{a}}$   & 94 & 14.5 & 0.4  \\ 
& 3HSP J180849.7+352042 $^{\text{a}}$  & 31 & 15. & 0.142 \\ \hline
\setrow{\bfseries}IceCube-131204A & VOU J191651-151902 $^{\text{b}}$ & 166 & 13.0 & -- \\ \hline 
 \end{tabular}}
\label{trackstable}
\end{center}
\end{table*}

\begin{table*}
\ContinuedFloat
\begin{center}
\caption{continued} 
\resizebox{\textwidth}{!}{
\def\arraystretch{1.3}
\begin{tabular}{L{4.0cm}L{5.0cm}L{2.cm}L{2.cm}L{2.cm}}
\hline
IceCube Name & Source Name &  $S_{\rm 1.4 GHz}$ & \nup & Redshift\\
 &  & [mJy] & [Hz]  &      \\ \hline \hline
IceCube-131023A & & & &\\ \hline
IceCube-170312A & & & &  \\ \hline
\setrow{\bfseries}IceCube-100710A & 3HSP J203031.7+223439 $^{\text{a}}$  & 5 & 16.2 & -- \\
 & 3HSP J203057.1+193612 $^{\text{a}}$  & 57 & 15.8 & 0.27 \\ \hline
IceCube-190124A & & & &  \\ \hline
IceCube-110128A & & & & \\ \hline
\setrow{\bfseries}IceCube-150714A & 3HSP J213314.3+252859 $^{\text{c}}$    & 40 & 15.2 & 0.294 \\
& VOU J213253+261144 $^{\text{d}}$  & 211 & 12.0 & 0.8 \\ \hline
IceCube-150812A & & & &\\  \hline
IceCube-120807A & & & & \\ \hline
IceCube-101009A & & & &  \\   \hline
\setrow{\bfseries}IceCube-140114A & 5BZB J2227+0037 $^{\text{a}}$   & 102 & 14.5 & --  \\   
 & 5BZQ J2226+0052 $^{\text{a}}$  & 615 & 12.5 & 2.26  \\   
 & 3HSP J222329.5+010226 $^{\text{b}}$  & 7 & 15.5 & 0.51\\  \hline  
IceCube-190331A &  &  & & \\ \hline
IceCube-171106A &  &  & & \\  \hline
IceCube-140108A &  &  & & \\  \hline
IceCube-140203A & & & &  \\ \hline
\setrow{\bfseries}IceCube-160510A  & VOU J232625+011147 $^{\text{c}}$   & 204 & 14.0 & 0.53 \\  \hline
\setrow{\bfseries}IceCube-190104A  &  IC 5362 $^{\text{a}}$ &  90  & 14.5 & 0.03 \\
& VOU J235815-285341 $^{\text{b}}$   &  169 & 14.0 & --   \\ 
  & 3HSP J235034.3-300604 $^{\text{d}}$ &  39  & 15.7 & 0.2328  \\  \hline
\end{tabular}}
\label{trackstable}
\end{center}
\end{table*}

\subsection{Dissecting the regions around the IceCube high-energy neutrinos}
\label{sec:vou}
The cross-matching with catalogues of astronomical sources is a widely used traditional method. Despite its simplicity and power, the effectiveness of this technique strongly depends on the catalogues available, which often do not include all the desired information, do not reach the maximum sensitivity for a specific search, and cannot provide the full discovery potential offered instead by the steadily growing amount of multi-frequency data available on the web. 
For this reason we have searched for the possible presence of \gr\,blazars in the uncertainty regions of the sample of IceCube tracks using VOU-Blazars, an innovative tool developed within the Open Universe initiative \citep{openuniverse} that makes extensive use of the information content of a large number of on-line multi-frequency databases.

VOU-Blazars \citep{dissecting,vou-blazar} is a software package that has been specifically designed to find blazars and blazar candidates in relatively large areas of the sky (e.g. uncertainty regions typical of \gr\ sources or even much larger, such as those of astrophysical neutrinos) on the basis of the multi-frequency data obtained using the Virtual Observatory (VO) methods developed by the International Virtual Observatory Alliance\footnote{\url{http://www.ivoa.net}}. 

The software locates sources that are or could be blazars using data from the latest blazar catalogues \citep[e.g. 5BZCAT, 3HSP, 4LAC:][]{5bzcat,3hsp,4LAC} and many multi-frequency surveys covering most bands of the electromagnetic spectrum \citep[e.g. NVSS, PCNT, AllWISE, Pan-STARRS, SDSS, GALEX, RASS, XMM, Fermi-LAT, MAGIC, VERITAS; see][for a complete list]{vou-blazar}. If requested the tool builds the radio to VHE \gr\ SED of any candidate found combining data from $\sim$ 70 catalogues and spectral databases retrieved using the conesearch and specsearch VO  protocols \citep{conesearch,specsearch}. These data-sets include the OUSXB database \citep{Giommi_2019a}, which is a complete image analysis of all the blazars observed by the \swift-XRT during the first 14 years of operations, and the detailed spectral analysis of bright blazars observed by \swift-XRT \citep{Giommi_2015,Giommi_2019b} 
in both Photon Counting (PC) and Windowed Timing (WT) observing modes \citep[see][for details]{xrt}.

\begin{figure*}
\centering
\includegraphics[width=7.4cm]{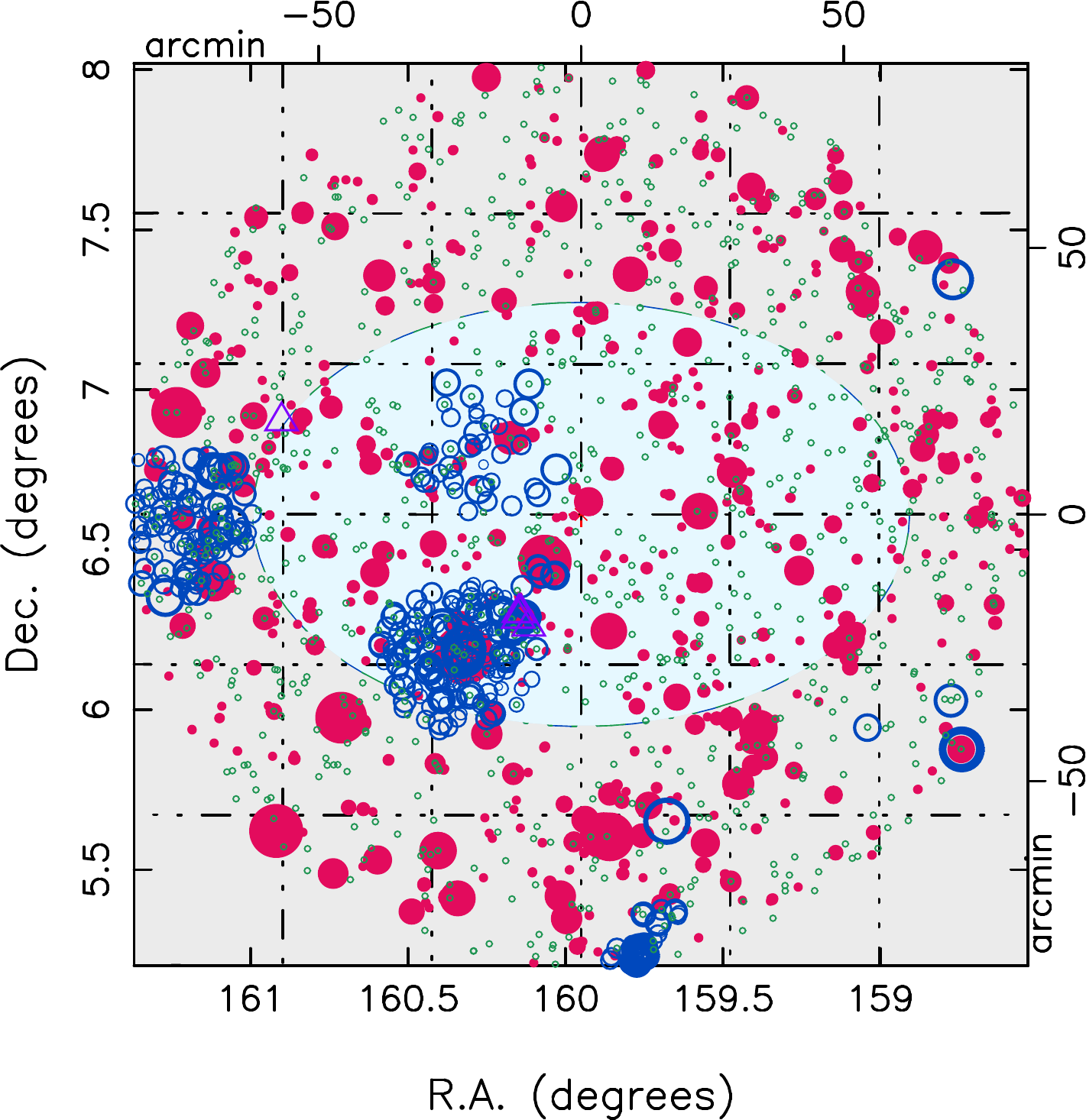}
\includegraphics[width=7.4cm]{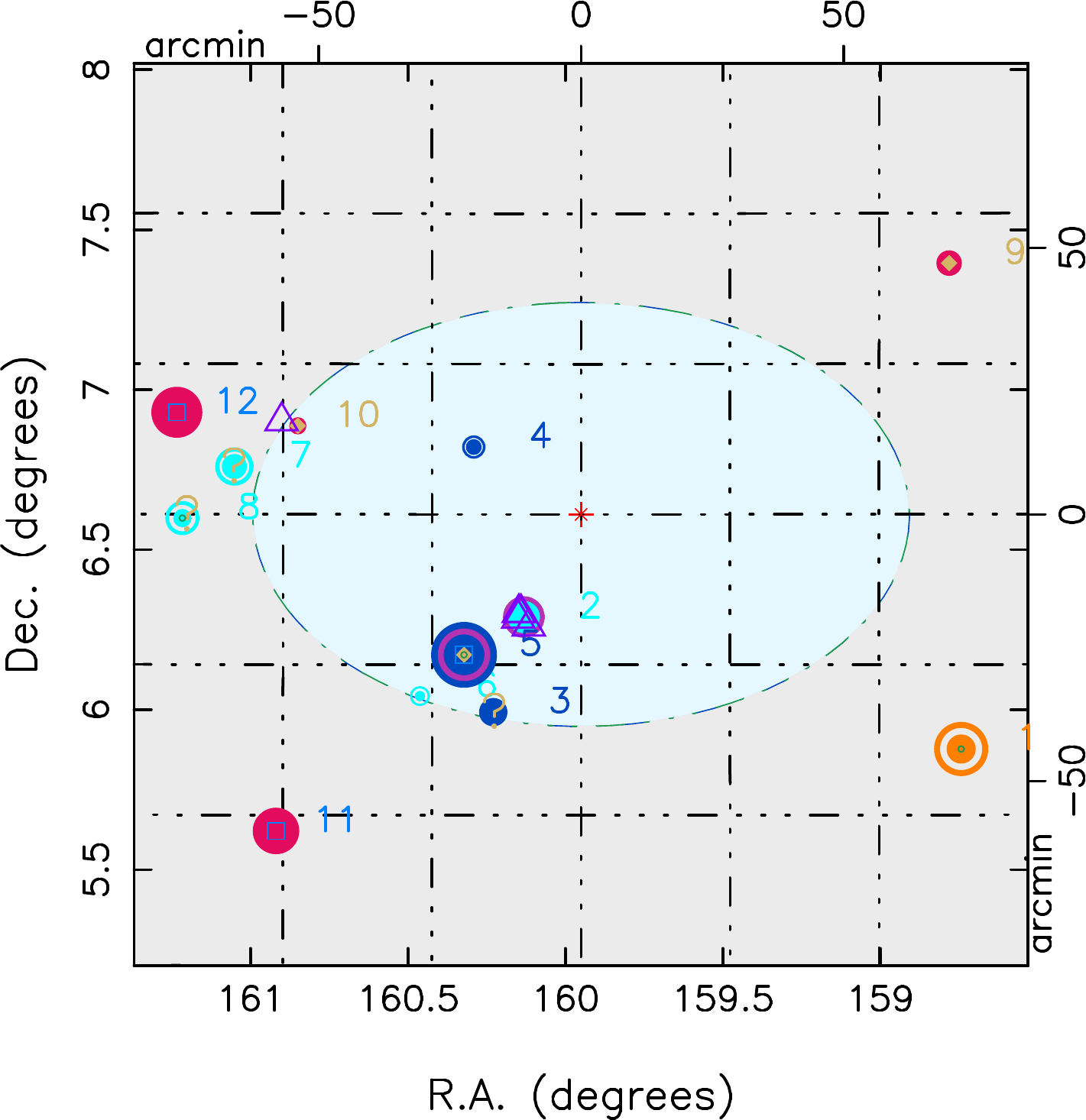} \includegraphics[width=2.7cm]{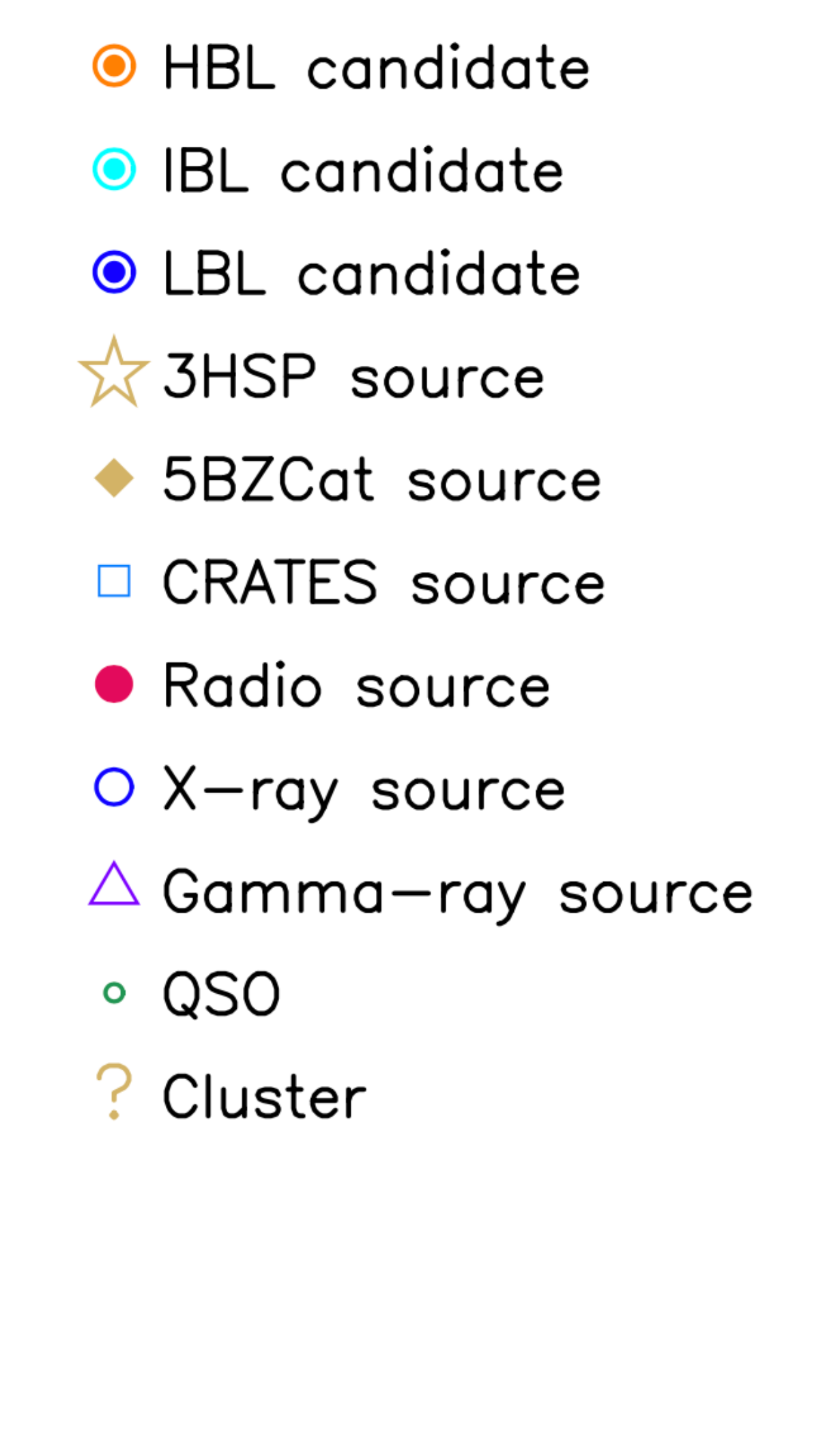}
\caption{This figure illustrates the complexity of the multi-frequency sky in fields as large as the uncertainty region of astrophysical neutrinos. The figure on the left plots the map showing all radio and X-ray sources in the FoV of IceCube-141209A.
Radio and X-ray sources are shown as red filled, and open blue circles circles respectively, with radius that is proportional to their flux density. The small green symbols represents known (mostly radio quiet) AGN in the field.
The figure on the right is the map showing blazar candidates in the field shown on the left. The radius of the filled circles is proportional to radio flux density, that of open circles is proportional to X-ray flux. 
The central elliptical region approximates the $\Omega_{90}$ of IceCube-141209A.}
\label{fig:vou-rxmap}
\end{figure*}

As an example of how VOU-Blazars works Fig. \ref{fig:vou-rxmap} shows the radio (red filled 
circles), X-ray (open blue circles), \gr\, (open triangles) sources, as well as known AGN 
from the 2019 edition of the Million Quasars catalogue \citep[small green 
circles:][]{milliquas} in a region covering the arrival direction of the neutrino 
IceCube-141209A.
The central elliptical area approximates the 90 per cent uncertainty region of this neutrino event ($\Omega_{90}$), and the  radius of the symbols is proportional to source intensity. Note that, while the AGN and the radio sources (from the NVSS survey: \citealt{NVSS}) are distributed in a rather uniform way, the density of X-ray sources is very uneven, reflecting the combination of data from the only full sky survey currently available in the X-ray band \citep[the Rosat All Sky Survey:][]{RASS_1999,RASS_2000} that is rather shallow, and observations performed with sensitive X-ray observatories like e.g. XMM, Chandra, or \swift\, where the density of X-ray sources detected within their small FoV is much larger.

The possible blazars found by the VOU-Blazars tool, based on spatial coincidence between radio and X-ray source and on their flux ratio, are plotted on the right side of Fig. \ref{fig:vou-rxmap}.
Different types of candidate blazars are shown with different colours: orange for objects with radio to X-ray flux ratio in the range observed for HBL blazars, cyan for IBL blazars and blue for LBL sources, 
 \citep[for a description of all the other symbols see the legend on the right side, or see][for more details]{vou-blazar}. Inside $\Omega_{90}$ of this IceCube neutrino three blazars are found, two of which are associated with \gr\, emission (marked with purple open triangles): source n. 5 is the FSRQ 5BZQ J1041+0610, source nr. 10 is the \gr\ detected BL Lac 5BZB J1043+0653 and source nr. 2 is GB6 J1040+0617 a previously uncatalogued IBL blazar associated with the \gr\, source 4FGL J1040.5+0617, that we designate VOU J104031+061721
 \citep[see also][]{Garrappa2019b}.

Once a candidate blazar is preliminarily identified based on the level of radio and X-ray emission, the VOU-Blazars tool can be requested to retrieve data from several other multi-wavelength catalogues and spectral databases and build the SED of the object for visual inspection.
For each candidate we carefully checked that its SED is fully consistent with that of a blazar, and we classified it as LBL, IBL or HBL, according to the location of \nup.
 The estimation of the latter is performed in two steps: 1. the data that can be attributed to components not related to the synchrotron emission from the jet, that is the host galaxy (usually in the IR), the blue bump  (Blue+UV), and inverse Compton emission (X-ray) are removed from the SED; 2. the remaining data are fitted with a two-degree polynomial using the SSDC-SED tool. If multiple observations of the object are available the fit is sensitive to the average flux.
When necessary, additional detailed investigation on the nature of each candidate was carried out using the Open Universe portal\footnote{\url{http://openuniverse.asi.it}}  \citep{openuniverse}, which provides facilitated access to a very large amount of data and services, the most widely used for this work being the SSDC SED builder\footnote{\url{https://tools.ssdc.asi.it/SED}}, the many optical, IR, and UV surveys available within Aladin light\footnote{\url{http://aladin.u-strasbg.fr/AladinLite}}, the 
SDSS\footnote{\url{http://skyserver.sdss.org/dr15}} and the DESI legacy surveys\footnote{\url{http://legacysurvey.org}} viewers, the ESA Sky tool\footnote{\url{http://sky.esa.int}}, the SSDC archive\footnote{\url{http://www.ssdc.asi.it/mma.html}}, 
the Vizier catalogues search tool\footnote{\url{http://vizier.u-strasbg.fr}}, the ESO science 
portal\footnote{\url{http://archive.eso.org/scienceporta}}, the bibliography services of ADS\footnote{\url{https://ui.adsabs.harvard.edu/}} and ArXiv\footnote{\url{https://arxiv.org/}}, as well as the NED\footnote{\url{http://ned.ipac.caltech.edu}} and 
CDS\footnote{\url{http://cdsportal.u-strasbg.fr}} portals.   \\

\subsubsection{Data analysis}
We used the VOU-Blazars tool to search for \gr\ detected blazars in all neutrino uncertainty areas and the regions immediately surrounding them.
To take into account that 10 per cent of the sources, by definition, are expected to lie outside the 90 per cent containment area ($\Omega_{90}$) and 
possible systematic uncertainties in the directional reconstruction, we have carried out our search in areas centred on the neutrino arrival directions and with size equal to $\Omega_{90}$ and then expanded by factors of 1.1, 1.3 and 1.5 ($\Omega_{90\times1.1},\Omega_{90\times1.3}, \Omega_{90\times1.5}$).

The sample of IceCube neutrino tracks and the results of our search for possible counterparts are given in Table\,\ref{sample} and Table\,\ref{trackstable}, respectively. In the latter the first column gives the IceCube event name, while the other columns give information about the possible counterpart candidates. Namely, column 2 gives the source name, column 3 the radio flux at 1.4 GHz, column 4 \nup, and column 5 the redshift as described in \ref{sec:redshift}.

Due to the difficulty in determining the exact \nup\ value when using non-simultaneous and sometimes sparse multi-frequency data of sources that are highly variable sources near their \nup\,, we grouped IBL and HBL blazars into a single category.

\subsubsection{Expectations from random coincidences}
To estimate the expected number of blazars in IceCube tracks due to random coincidences we have carried out exactly the same procedure described above in a control area composed of circular regions with $3^{\circ}$ radii each, centred on the same Right Ascension of each detected neutrino, and with declination increased or decreased by a fixed amount of $6^{\circ}$.  To reproduce the same conditions of the statistical sample, circles that after the shift in declination had a position within ten degrees of the Galactic plane were not used. This procedure led to a control sample covering a total area of 2,573 square degrees. 

The same VOU-Blazars procedure used for the statistical sample, applied to the control sample, led to the detection of 103 \gr\ blazars of the LBL type and 103 \gr\ blazars of the IBL/HBL type, leading to an expected average density that is identical for the two types of blazars of one object every 27.4 square degrees. 

\subsubsection{Statistical analysis methods}
\label{sec:stat_method}

To evaluate of the compatibility of the observed statistics with random expectations, as well as a quantitative estimation of the number of blazars that could be associated with IceCube neutrinos, we have applied two methods: 1. a direct comparison of the observed blazars counting with the expected number of random coincidences, with the corresponding probability calculated using Poisson statistics; 2. a likelihood ratio test, as described below.
The observables of the likelihood method include the total number of blazars and the distribution of observed matches, that is the number of neutrino regions with zero counterparts ($n_0$), one counterpart ($n_1$), two counterparts ($n_2$), etc. The final set of observables can thus be written as $\theta = (n_0, n_1, n_2, n_3, n_{\geq 4})$. We sum all the neutrino regions with 4 or more counterparts in one bin, as they are by construction very rare and do not carry significant information. The hypotheses tested are defined as follows:

\begin{itemize}
    \item \textbf{Background Hypothesis (H0)}: each neutrino uncertainty region has an associated number of expected background sources,  which depends on the average source density as estimated from the control region, and on the size of the error region.
    \item \textbf{Signal Hypothesis (H1)}: in addition to the background sources, there is a number $N_{srcs}$ of neutrinos with a signal counterpart.
\end{itemize}
The test statistic is defined as $\mathcal{TS} = -2\times\log \lambda$, with
\begin{equation}
    \lambda = \frac{\mathcal{L}_{H0}}{\mathcal{L}_{H1}}=\frac{p(n_1, n_2, n_3, n_{\geq 4}\,|\,\textrm{H0} )}{\textrm{max}_{N_{srcs}}\,\,p(n_1, n_2, n_3, n_{\geq 4}\,|\,\textrm{H1} )}
\end{equation}
and probability density functions (pdfs) $p(n_1, n_2, n_3, n_{\geq 4})$. The denominator maximises the signal pdf over the number of signal sources. Note that the pdfs do not depend anymore on $n_0$, as it is not a free parameter given the other observables. The pdfs are generated for the various error regions ($\Omega_{90\times1.0}$, $\Omega_{90\times1.1},\Omega_{90\times1.3}, \Omega_{90\times1.5}$) and different signal strength $N_{srcs} \in \{0,1,\dots, 40\}$ using a Monte Carlo simulation based on the sample of 70 neutrinos considered in this work. For each trial in the simulation we first draw a Poisson number of background sources for every region, before we, in a second step, randomly distribute $N_{srcs}$ signal sources uniquely over the neutrinos. Using the same trial generation method also the test-statistic distributions for the background and signal hypothesis are generated for the subsequent calculation of p-values and limits.

\subsection{Results}
\label{sec:results-VOU}

The search described in Sec.\,\ref{sec:stat_method} lead to the identification of a total of 72 \gr\, detected blazars in spatial coincidence with at least one error area associated with the IceCube events.
In the largest coverage areas $\Omega_{90\times1.5}$,
 we found 47 \gr-detected blazars with SED typical of IBL/HBL sources, one of the brightest radio galaxy/blazar in the sky (M87) and 24 objects with SEDs typical of LBL blazars.
Since the expectations from the control sample for the two types of blazars (LBLs and IBLs/HBLs) are identical (26.8), the large excess of $\sim$ 20 IBL/HBL sources ($47-26.8$) already points towards an overabundance of this type of sources, compared to the random sky.

The details of the statistics of the search and the comparison with the control sample are summarised in Table\,\ref{resultstable}. The rows show the results for the different error regions as given in column 1. Column 2, 3, 4 and 5, 6, 7 give the results for the IBL/HBL and LBL \gr\ sources, respectively. Additionally to the number of counterparts we also give the expectation from the control area as well as the result from the likelihood ratio test.

\begin{figure}
\centering
\includegraphics[width=1.\linewidth]{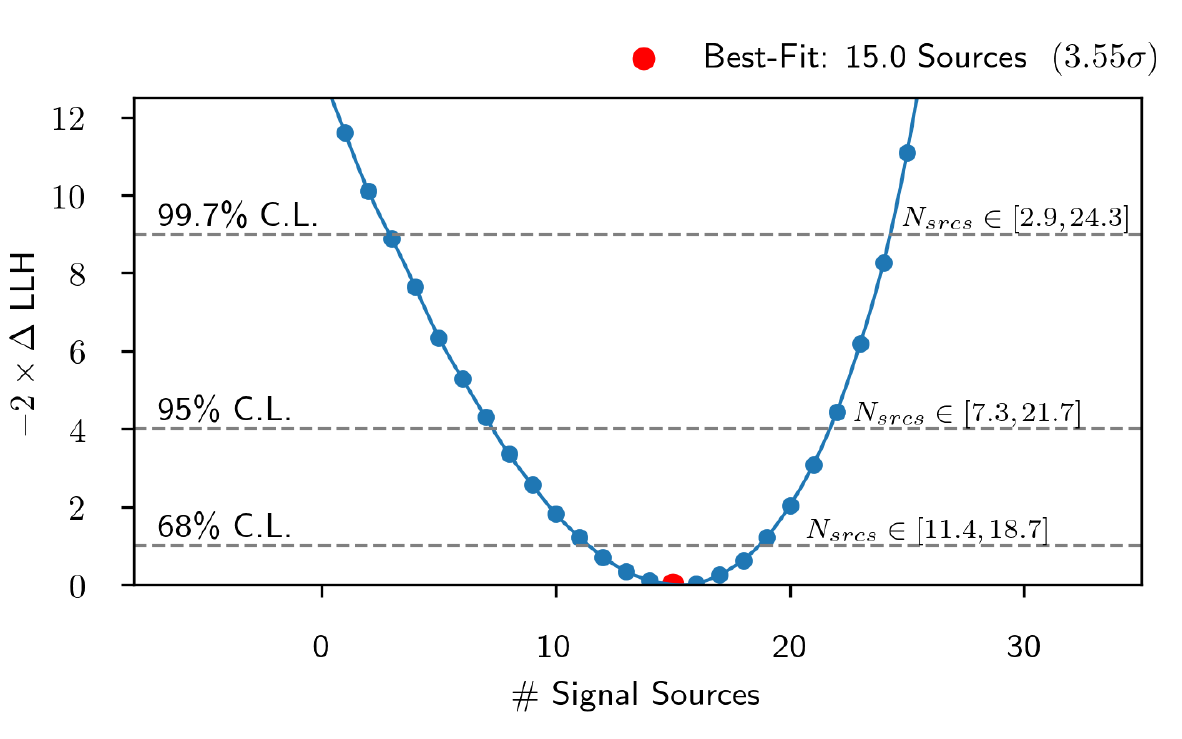}
\caption{The profile likelihood for the class of HBLs/IBLs in $\Omega_{90\times1.3}$. The best-fit is a number of $15 \pm 3.6$ signal sources at $1 \sigma$ confidence level. The background hypothesis is excluded at 3.56$\,\sigma$.}
\label{fig:llh_res}
\end{figure}

The smallest p-value in our test is obtained for the class of IBLs/HBLs in $\Omega_{90\times1.3}$. While only 20.1 sources are expected we actually observe 35 which is equivalent to a Poisson p-value of $1.5 \times 10^{-3}$. Using the full likelihood approach described in Sec.\,\ref{sec:stat_method} we find a test-statistic value of 12.51, which is - after comparison to the background test-statistic distribution - equivalent to a significance of $3.56\,\sigma$ with a best-fit number of $\hat{N}_{srcs} = 15 \pm 3.6$ ($\hat{N}_{srcs} = 16 \pm 4$ for  $\Omega_{90\times1.5}$). The corresponding profile likelihood is shown in Fig. \ref{fig:llh_res}. The confidence levels are calculated assuming a $\chi^2_1$ distribution of the test-statistics, which has been verified using Monte Carlo simulations. Since we have performed the same test for $\Omega_{90\times1.0}$, $\Omega_{90\times1.1},\Omega_{90\times1.3}, \Omega_{90\times1.5}$, we need to correct the p-value with the \textit{effective number} of independent trials. From Monte Carlo simulations we obtain a trial factor of 1.64. Given that we tested the two classes of IBLs/HBLs and LBLs separately, the overall trial factor becomes 3.28 giving a final post-trial p-value of $6.2 \times 10^{-4}$ (3.23 $\sigma$).

\begin{figure}
\centering
\includegraphics[width=1.\linewidth]{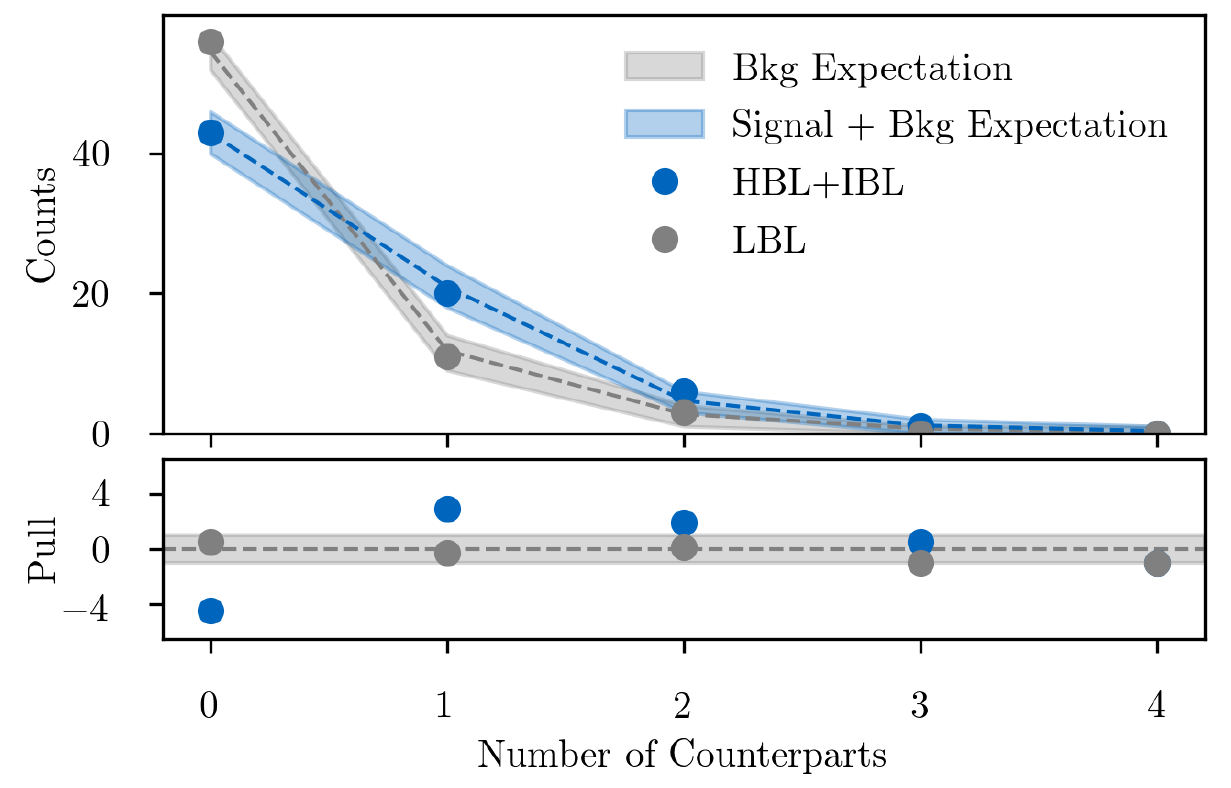}
\caption{Top panel: Distribution of the number of counterparts for HBLs/IBLs and LBLs in $\Omega_{90\times1.3}$. The grey and blue shaded bands show the expected mean and standard deviation for the case of pure background and background plus 15 signal sources as calculated from Monte Carlo. The dots show the experimental results. In the bottom panel the deviation of the experimental data from the background is shown in terms of Gaussian $\sigma$.}
\label{fig:gof}
\end{figure}

To evaluate if the observed excess of counterparts is actually consistent with a signal expectation, we have compared the distributions of the number of sources identified in $\Omega_{90\times1.3}$ with the expectation from background and the best-fit signal of $\hat{N}_{srcs} = 15$ sources, as shown in Fig.~\ref{fig:gof}. The shaded bands in the top panel show the expected mean and standard deviation estimated from a Monte Carlo simulation. It can be seen that, while the class of LBL is very much consistent with the background expectation, there is a clear deviation for the class of IBLs/HBLs. 
The bottom panel shows the deviation of the data from the background expectation in terms of Gaussian $\sigma$. The largest contribution to the excess of IBLs/HBLs comes from the bin with a number of counterparts equal to 1; while we expect 11.8 sources, we observe 20 objects instead.

\begin{figure}
\centering
\includegraphics[width=1.\linewidth]{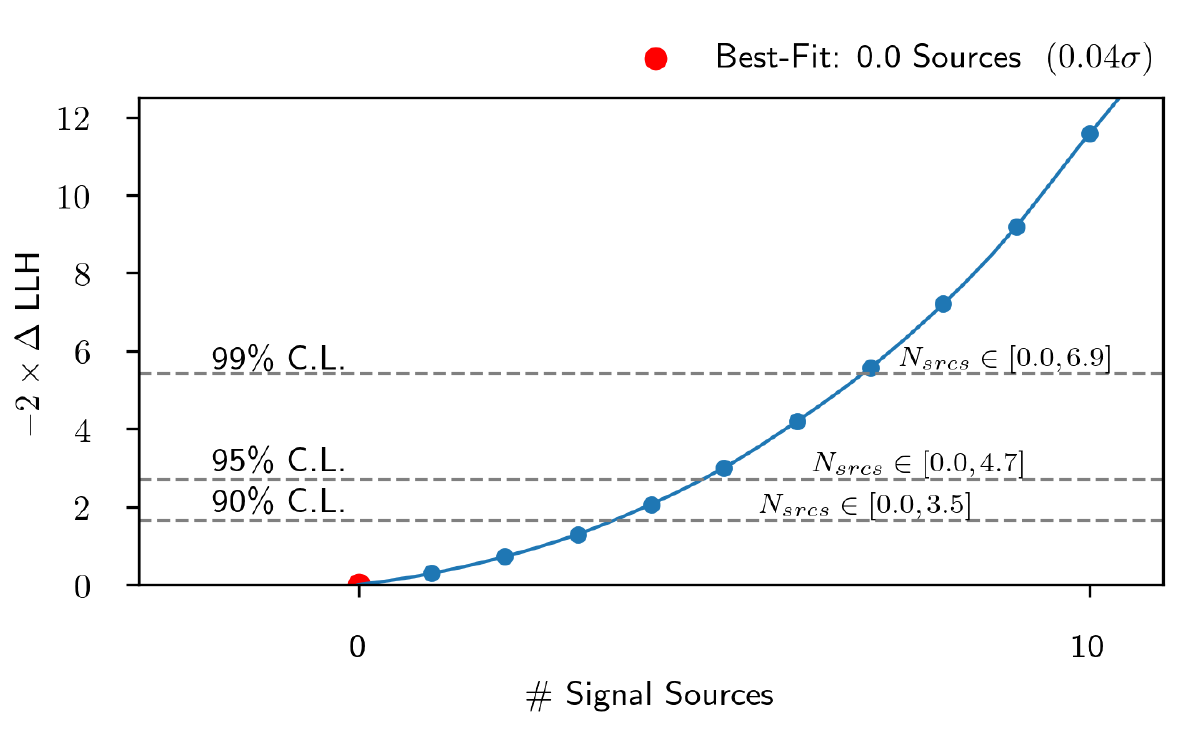}
\caption{The profile likelihood for the class of LBLs in $\Omega_{90\times1.3}$. As there is no significant detection we place an upper limit of $N_{srcs}^{UL} = 3.48$ at 90\% C.L.}
\label{fig:llh_res_lbl}
\end{figure}

For LBLs on the contrary we do not observe any excess over the background, which allows us to place an upper limit on the maximum number of signal LBLs in our sample. The corresponding profile likelihood for $\Omega_{90\times1.3}$ is shown in Fig. \ref{fig:llh_res_lbl}. Using again the $\chi^2_1$ distribution of the test-statistic we get an upper limit of 3.48 sources at 90\% C.L.

\subsubsection{Redshift distribution}
\label{sec:redshift}

\begin{figure}
\centering
\includegraphics[width=1.\linewidth]{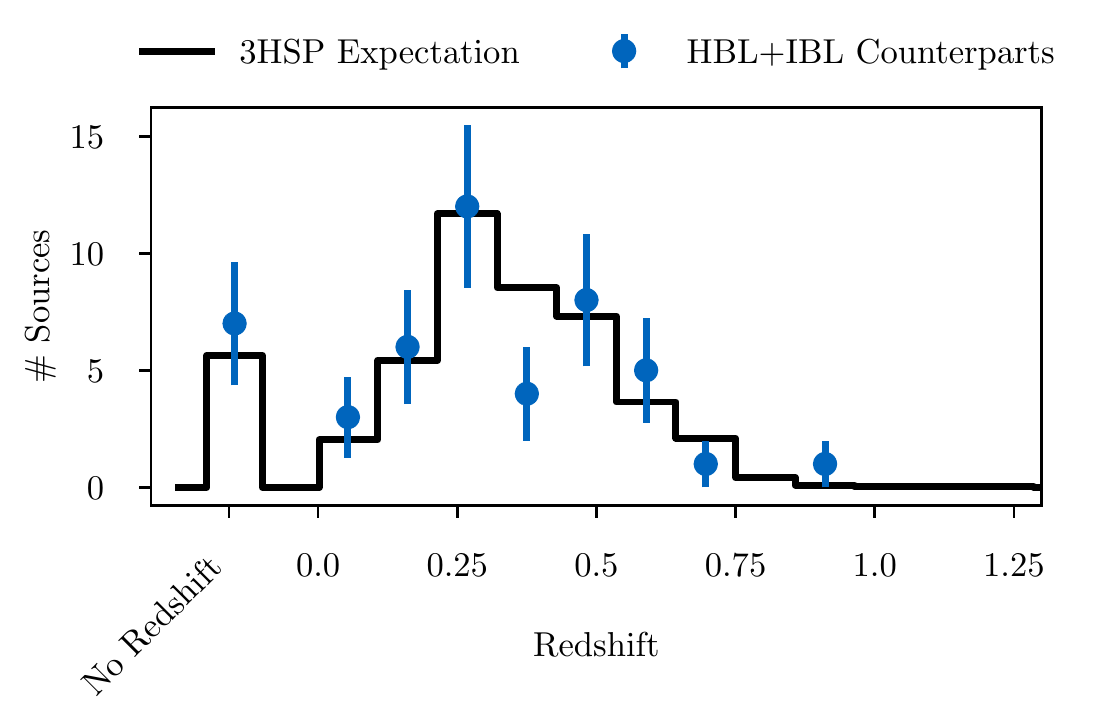}
\caption{Comparison of the redshift distribution of IBL/HBL blazars found in our study and the 3HSP sample. }
\label{fig:hsp_redshift}
\end{figure}

The redshifts of the blazars listed in Table\,\ref{trackstable} have been taken from the original catalogues, (e.g. 5BBZCAT, 3HSP or  4FGL), or from the on-line ZBLLAC\footnote{https://web.oapd.inaf.it/zbllac/} database. For the case of the objects denoted with the VOUJ prefix, that is blazars that were not previously reported in the literature and were found in our search using the VOU-BLazars tool, the redshift has been estimated using the photometric redshift estimation described in \cite{3hsp}. 
The redshift distribution of the sample of 47 IBL/HBL sources is shown in Fig. \ref{fig:hsp_redshift}, compared to the redshift distribution of the sources in the 3HSP catalogue (solid histogram). Excluding the sources for which no redshift could be estimated the redshift ranges from 0.0043 to 0.95 with a median value of 0.32. The fraction of sources with no redshift (8 out of 47 objects) is slightly higher than that in the 3HSP catalogue. However, several sources are of the IBL type and it is difficult to compare this fraction with the background expectations as no reliable samples of IBL sources is available.

\begin{table*}
\begin{center}
\caption{Results on the occurrence of \gr\ blazars within the 70 IceCube high Galactic latitudes ($|b|>10^{\circ}$) neutrino with error radii $\le 3.0^{\circ}$ and comparison with the expectations
due to random coincidences as estimated from the control sample.}
\begin{tabular}{l|ccc|ccc}
\hline
Area searched &\gr\ IBL/HBL  & Expectation & Likelihood-test  & \gr\ LBL  & Expectation & Likelihood-test    \\
             & found in neutrino & from control & p-value & found in neutrino & from control & p-value  \\
             & error region & sample &                          &  error region &  sample &  \\
(1) & (2) & (3) & (4) & (5) & (6) & (7) \\
\hline\hline
$\Omega_{90}$ & 20 & 11.9 &$7.4\times10^{-3}$ & 9 & 11.9 & 0.43 \\
$\Omega_{90\times1.1}$ & 24 & 14.4 & $1.4\times10^{-2}$ & 15 & 14.4 & 0.44 \\
$\Omega_{90\times1.3}$ & 35 & 20.1 & $1.9\times10^{-4}$ & 17 & 20.1 & 0.48 \\
$\Omega_{90\times1.5}$ & 47 & 26.8 & $2.0\times10^{-4}$ & 24 & 26.8 & 0.33 \\
\midrule
\multicolumn{6}{c}{Post trial p-value: $6.2 \times 10^{-4}$ (3.23 $\sigma$)}\\
\midrule
\bottomrule
\end{tabular}
\label{resultstable}
\end{center}
\end{table*}

\section{Neutrinos and gamma-ray sources not included in the statistical sample}

\subsection{M87}
The giant radio galaxy M87, one of the brightest and most remarkable objects in the extragalactic sky, is within the uncertainty region of the IceCube neutrino IceCube-141126A.
Despite this object being normally referred to as a classical radio galaxy, the jet inclination angle of only $\sim 17^{\circ}$ \citep{Walker_2018} and the superluminal motion that has been detected at radio, optical and X-ray frequencies \citep{Cheung,Biretta,Snios} make it {\it almost} a blazar \citep{UP95}. 
Moreover, similarly to HBLs, M87 is a strong emitter of \gr s in the GeV and VHE band, 
where it shows a flat spectrum and large flux variability \citep{4FGL,Bangale}. Due to 
the complexity of the optical and radio emission of this source it is not possible to 
estimate a reliable value of \nup\,, and for this reason we have not 
included it in any of the samples of Tab. \ref{resultstable}.

\begin{figure*}
\centering
\includegraphics[width=17.5cm]{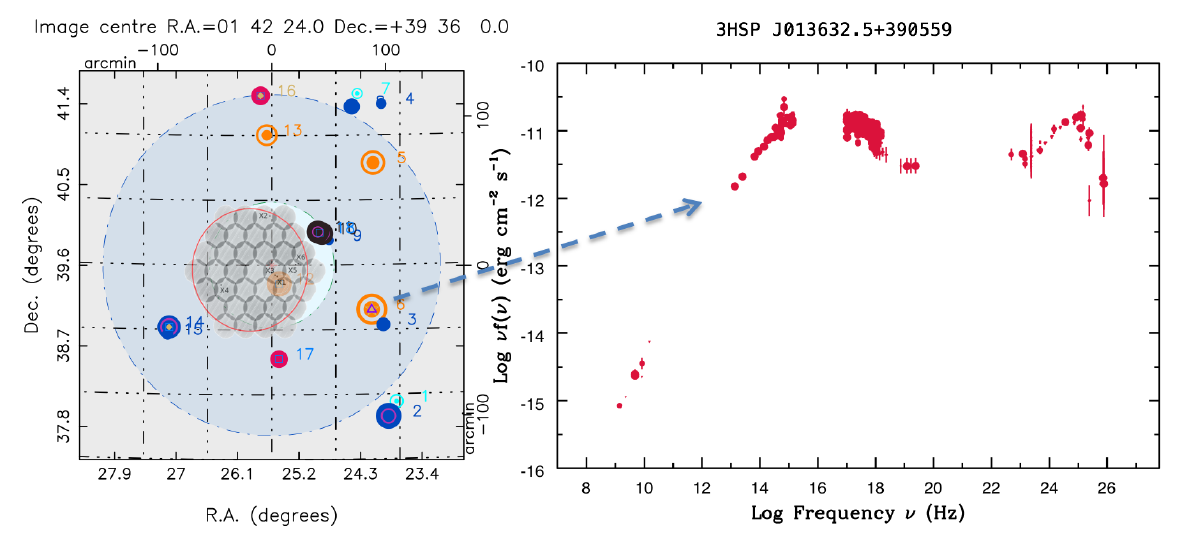}
\caption{The region around the neutrino triplet event and the SED of the \gr\ detected blazar 3HSPJ013632.6390559. The gray structure inside the red circle \protect\citep[adapted from][]{threeneutrinos} represents the 37 tiling observations that have been performed by \swift\ shortly after the detection of the three neutrinos covering the 50 per cent error region centred on the initially published arrival direction. The light and darker blue circular regions represent the revised 50\% and 90\% error regions respectively.
}
\label{fig:3neutrinos}
\end{figure*}

\begin{figure}
\centering
\includegraphics[width=8.3cm]{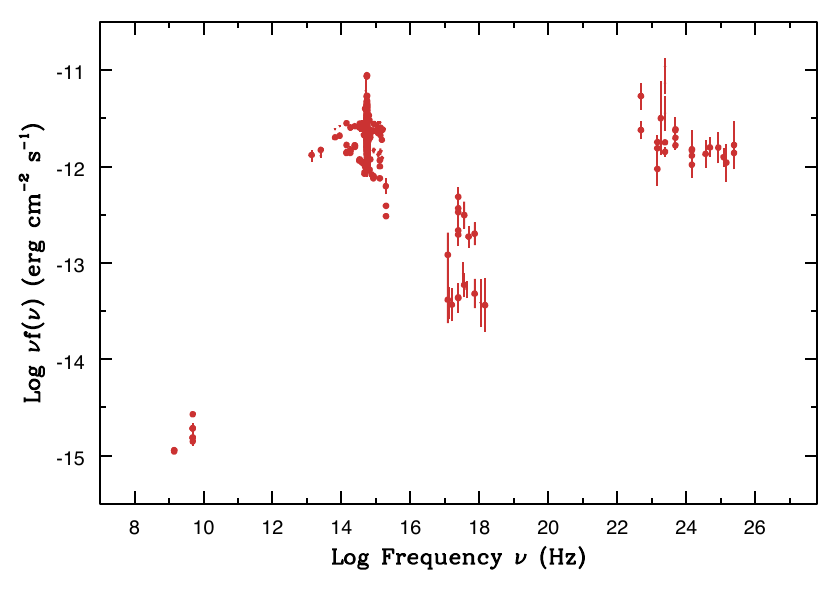}
\caption{The SED of MG3J225517+2409, the blazar that could be the counterpart of five ANTARES neutrinos and one IceCube track event with
relatively large positional uncertainty. The shape of the SED shows that this object is of the IBL type, similar to TXS\,0506+056.}
\label{fig:antares}
\end{figure}

\subsection{The case of the neutrino triplet}

In February 2016 the IceCube observatory detected three lower energetic neutrinos ($\sim$1\,TeV) that arrived from directions consistent with a single source and within 100\,s of each other \citep{threeneutrinos}.
This triple neutrino detection was considered to be very unlikely due to a random coincidence and therefore its announcement triggered a number of multi-wavelength follow-up observations. In particular, \swift\ mapped the 50 per cent uncertainty area with a series of 37 short XRT tiled observations. The left side of Fig. \ref{fig:3neutrinos} shows the 37 \swift\,-XRT X-ray images in grey colour scale, inside a red circle. The light blue central area, largely overlapping the XRT pointings, is the 50 per cent confidence region, recalculated with a different algorithm. 
The larger blue circle represents the 90 per cent error region of the neutrino triplet, too large to be covered with a reasonable number of \swift\ observations. No likely counterpart was found in the 50 per cent containment area covered by the \swift\ data. However, the VOU-Blazars map shown in the left part of Fig. \ref{fig:3neutrinos} reveals the presence of some blazars, one of which is the \gr\ detected HBL blazar 3HSPJ013632.5+390559 whose SED is plotted on the right-side of Fig. \ref{fig:3neutrinos}. 3HSPJ013632.5+390559 was not observed by \swift\ as it is outside the 50 per cent error region, and therefore we do not know if it was flaring in the X-rays in that period.

\subsection{IceCube-100608A and ANTARES neutrinos near the blazar MG3~J225517+2409}

The IceCube event IceCube-100608A is not included in the sample of 70 tracks considered above because the uncertainty in its arrival direction is $> 3^{\circ}$. Nevertheless, we discuss it in this section since its error region includes the blazar MG3 J225517+2409, which has been recently reported as a possible counterpart of 5 ANTARES track events with energy ranging form $\sim 3$ to $\sim 40$ TeV \citep{antares2019}. The SED of MG3 J225517+2409, assembled using the latest multi-frequency information available, plotted in Fig. \ref{fig:antares} shows that this object is an IBL blazar similar to TXS\,0506+056, 
and not an LBL object as initially reported in \cite{antares2019}. \cite{Paiano_2019} have derived a lower limit of z $> 0.863$ to the redshift of this source. 

\subsection{IceCube-190704A and the blazar 3HSPJ104516.3+275133}

This event has a localisation error $>3^{\circ}$. It is however worth discussing it here since it includes the blazar 3HSP J104516.3+275133 (=1WHSP J104516.2+275133), which, although not listed in any \fermi\,-LAT catalogue, has been reported by the \fermi\ team as a \gr\ detected source in a telegram \citep{Garrappa2019a} that was published shortly after the announcement of IceCube-190704A. 

The detection of 3HSP J104516.3+275133, a previously unknown \gr\, blazar, reported by the \fermi-LAT team following a specific analysis using all the data collected until then, suggests that other \gr\, emitting blazars not listed in the \fermi\, 4FGL catalogue might be present in the error regions of IceCube neutrinos. 
To find how many such sources are present we plan to carry out a detailed analysis of all neutrino events using the complete data-set available in the \fermi\,-LAT archive ($\gsim 11$ years) and applying the same statistical significance threshold as done by the \fermi-LAT team. The results will be presented in future publications. 

\section{Discussion and conclusions}\label{sec:conclusion}

In an effort to test the association between astrophysical neutrinos and blazars using a large statistical sample, we have compiled a list of 94 high-energy IceCube neutrinos that have been published as 
through-going tracks, high-energy starting tracks or alerts in the IceCube's realtime program.
Using the sub-sample of 70 events detected above the Galactic plane (|b|$>$10$^{\circ}$) with positional uncertainty $< 3^{\circ}$ we searched for possible excesses of \gr\,blazars following two approaches:
1) a cross-matching with catalogues of \gr\,sources and \gr\,emitting blazars; 2) a careful dissection analysis of each IceCube track, similar to that presented by \cite{dissecting} on the association of the blazar TXS\,0506+056 with the IceCube-170922A neutrino, using the VOU-Blazars tool and the services available within the Open Universe portal. 

Our main results can be summarised as follows:
\begin{itemize}
\item A 2.79\,$\sigma$ excess was found using the method of positional cross-matching with the 3HSP catalogue of high energy peaked blazars. 
No significant excesses were instead found in the cases of the \fermi-4LAC and \fermi-3FHL catalogues (see Table\,\ref{statantable} for details).

\item The dissection analysis based on the VOU-Blazars tool shows an excess of IBL/HBL \gr\ detected blazars in correspondence to IceCube neutrino positions. 
The most significant result is obtained for the case of a search in 1.3 times the 90\% error region ($\Omega_{90\times1.3}$), with a post trial p-value of $\sim 6.2\times 10^{-4}$, corresponding to 3.23\,$\sigma$, and an excess of approximately 15 IBL/HBL blazars compared to the expectation.
The fact that the most significant excess  obtained is within $\Omega_{90\times1.3}$ and not within $\Omega_{90}$ may indicate that a broader coverage of the angular uncertainties of IceCube is required for follow up searches, and might be the symptom of the presence of some level of systematic uncertainty in the estimation of the arrival direction of IceCube neutrinos.

\item No excess is found for the case of \gr\ detected LBL blazars.

\item The very bright radio galaxy M87 is inside the error region of one IceCube neutrino. This is the only object of this type in the part of the sky accessible to high-energy neutrinos by IceCube ($-35 ^{\circ}< \delta < +35^{\circ}$, $\sim$ 23,600 square degrees, see Fig. \ref{sample}) and the probability to be included by chance in one of the 70 tracks considered, which cover a total area of $\sim$ 290 square degrees, is 290/23,600, or $\sim$ 0.012. The only other similar source, Centaurus A is located outside the area with declinations between $-35^{\circ}$ and +35$^{\circ}$. 

\end{itemize}

 Our study shows that there is an excess of $\sim$  15\,\gr\, IBL/HBL blazars in the sample of 70 IceCube neutrino events. The statistical significance of this result ($>3\sigma$) together with previous works \citep[e.g.][]{Pad_2014,Resconi_2017,Lucarelli_2017,Lucarelli_2019,iconly,antares2019} show that there is a persistent and growing evidence that IBL/HBL sources are the most likely counterparts of a fraction of IceCube's high-energy neutrinos.

If we assume that the excess is not a rare  statistical fluctuation but reflects a true association between blazars and cosmic neutrinos, this would correspond to one \gr\ IBL/HBL blazar every $\sim$ 4.5 neutrinos, or $\sim$\,21 per cent.
Considering that the IceCube tracks have some probability of not being of astrophysical origin, the fraction of IceCube astrophysical neutrinos that could be associated with blazars could be even higher. This is not in tension with the non detection from IceCube stacking of $\gamma$-ray blazars (\citealt{ICECube17_1}, \citealt{Huber:2019lrm}). As shown by Table\,1 of the latter paper, the stacking upper limit in fact strongly depends on the assumed neutrino spectrum. Note additionally that our result is not based on any pre-existing catalogue but on a detailed multi-wavelength study of the regions associated with the IceCube neutrinos.

The list of IBL/HBL blazars that have been found within 1.5 times the 90\% error region is given in Table\,\ref{trackstable}. This sample of 47 objects has a best-fit of 16 signal counterparts. We have therefore looked for possible differences or peculiarities with respect to other samples. Fig.~\ref{fig:hsp_redshift} compares the redshift distribution of this sample and of that of the 3HSP catalogue. No significant differences are present. The only deviation worth mentioning is a slight overabundance of objects with no redshift. Similar conclusions can be made for the case of the radio flux density distribution. A much more detailed multi-frequency and time-domain analysis
of this sample will be presented in a future paper.


Today's major limitations are the still small number of neutrino events detected and 
the large uncertainty in the arrival directions \citep[e.g.][]{Pad_2014,dissecting}. Using the Monte Carlo simulation described in Section \ref{sec:stat_method} we find that the significance $\Sigma$ of our analysis is expected to scale with time $T$ as $\Sigma \propto T^{0.55}$, consistently with the fact that we are working in a regime of dominant Poisson statistics. We have therefore estimated that in order to achieve a $5\sigma$-level association between astrophysical neutrinos and blazars, the neutrino statistics will have to more than double. Future neutrino telescopes such as Baikal-GVD\footnote{\url{https://baikalgvd.jinr.ru}}, IceCube-Gen2\footnote{\url{https://icecube.wisc.edu}}, KM3NeT\footnote{\url{https://www.km3net.org}}, and P-ONE\footnote{\url{http://www.pacific-neutrino.org}} will be then crucial to substantially contribute to a rapid improvement on neutrino statistics over the whole sky.
The steady progress in the field, together with possible future improvements in the localisation of IceCube tracks, could then turn the mounting evidence described above into a conclusive statistical result, settling the long-debated question of the association of cosmic sources with high-energy astrophysical neutrinos in favour of the type of blazars whose electromagnetic emission reaches the largest observed energies.

\section*{Acknowledgments}

We acknowledge the use of data from the IceCube public archive and several astronomical archives, obtained with the software and services of the United Nations' ``Open Universe'' initiative, the Space Science Data Center, SSDC managed by the Italian Space Agency, and the International Virtual Observatory Alliance.
PP thanks the SSDC for the hospitality and partial financial support for his visit. 
PG and TG acknowledge the support of the Technische Universit{\"a}t M{\"u}nchen -
Institute for Advanced Study, funded by the German
Excellence Initiative (and the European Union Seventh Framework
Programme under grant agreement n. 291763).
This work is supported by the Deutsche Forschungsgemeinschaft through grant SFB\,1258
``Neutrinos and Dark Matter in Astro- and Particle Physics''.

\label{lastpage}

\bsp	


\begin{thebibliography}{}
\bibitem[\protect\citeauthoryear{Aartsen et al.}{2013}]{2013PhRvL.111b1103A} Aartsen M.~G., et al., 2013,
 Phys. Rev. Let., 111, 021103
\bibitem[\protect\citeauthoryear{Aartsen et al.}{2014}]{Aartsen2014} Aartsen M.~G., et al., 2014, PhRvL, 113, 101101 
\bibitem[\protect\citeauthoryear{Aartsen et al.}{2015}]{Aartsen2015}
  Aartsen M.~G., et al., 2015, Phys. Rev. Let., 115, 081102
\bibitem[\protect\citeauthoryear{Aartsen et al.}{2016}]{Aartsen2016}  Aartsen M.~G., et al., 2016, ApJ, 833, 3
\bibitem[\protect\citeauthoryear{Aartsen et al.}{2017}]{Aartsen2017}
  Aartsen M.~G., et al., 2017, ApJ, 835, 45
\bibitem[\protect\citeauthoryear{Acero et al.}{2015}]{Fermi3FGL}
Acero F., et~al., 2015, \apjs, 218, 23
\bibitem[\protect\citeauthoryear{Ackermann et al.}{2011}]{Fermi2LAC}
  Ackermann M., et al., 2011, ApJ, 743, 171
\bibitem[\protect\citeauthoryear{Ackermann et al.}{2015}]{Fermi3LAC}
  Ackermann M., et al., 2015, ApJ, 810, 14
\bibitem[\protect\citeauthoryear{Ackermann et al.}{2016}]{2FHL} Ackermann
  M., et al., 2016, ApJS, 222, 5
\bibitem[\protect\citeauthoryear{Ajello et al.}{2017}]{3FHL} 
Ajello M., et al., 2017, ApJS, 232, 18  
\bibitem[\protect\citeauthoryear{Aublin}{2019}]{antares2019} Aublin J., on behalf of the ANTARES Collaboration, 
2019, 36th International Cosmic Ray Conference, 840 
\bibitem[\protect\citeauthoryear{Bangale et al.}{2015}]{Bangale} Bangale P., Manganaro M., Schultz C., Colin P., Mazin D., 2015, ICRC,  759, ICRC...34
\bibitem[\protect\citeauthoryear{Biretta, Sparks \& Macchetto}{1999}]{Biretta} Biretta J.~A., Sparks W.~B., Macchetto F., 1999, ApJ, 520, 621
\bibitem[\protect\citeauthoryear{Brown, Adams, \& Chadwick}{2015}]{Brown_2015} 
Brown A.~M., Adams J., Chadwick P.~M., 2015, MNRAS, 451, 323
\bibitem[\protect\citeauthoryear{Burrows et al.}{2005}]{xrt} Burrows D.~N., et al., 2005, SSRv, 120, 165
\bibitem[\protect\citeauthoryear{Chang et al.}{2017}]{2whsp}
Chang Y.L., Arsioli B., Giommi P. \& Padovani P., 2017, A\&A, 598, 17
\bibitem[\protect\citeauthoryear{Chang et al.}{2019}]{3hsp}
Chang Y.L., et al., 2019, A\&A, 632, A77
\bibitem[\protect\citeauthoryear{Chang, Brandt, \& Giommi }{2020}]{vou-blazar}
Chang Y.L., Brandt C., \& Giommi P., 2020, Astronomy \& Computing, 30, 100350
\bibitem[\protect\citeauthoryear{Cheung, Harris \& Stawarz}{2007}]{Cheung} Cheung C.~C., Harris D.~E., Stawarz {\L}., 2007, ApJL, 663, L65
\bibitem[\protect\citeauthoryear{Condon et al.}{1998}]{NVSS}
Condon J.~J., Cotton W.~D., Greisen E.~W., Yin Q.~F., Perley R.~A., Taylor G.~B.,
Broderick J.~J., 1998, AJ, 115, 1693
\bibitem[\protect\citeauthoryear{Fermi-LAT collaboration}{2019a}]{4FGL} Fermi-LAT collaboration, 2019a, arXiv:1902.10045 
\bibitem[\protect\citeauthoryear{Fermi-LAT collaboration}{2019b}]{4LAC}
Fermi-LAT collaboration, 2019b, arXiv:1905.10771
\bibitem[\protect\citeauthoryear{Flesch}{2017}]{milliquas} Flesch E.~W., 2017, yCat, VII/280
\bibitem[\protect\citeauthoryear{Garrappa, Buson \& Venters}{2019a}]{Garrappa2019a} Garrappa S., Buson S., Venters T., 2019a, ATel, 12906, 1
\bibitem[\protect\citeauthoryear{Garrappa, et al.}{2019b}]{Garrappa2019b} Garrappa S., et al., 2019b, ApJ, 880, 103
\bibitem[\protect\citeauthoryear{Geherls et al.}{2004}]{swift}
Gehrels N. et al. 2004, ApJ, 611, 1005
\bibitem[\protect\citeauthoryear{Ghisellini et al.}{2017}]{Ghisellini_2017}
Ghisellini G., Righi C., Costamante L., Tavecchio F., 2017, MNRAS, 469, 255
\bibitem[\protect\citeauthoryear{Giommi}{2015}]{Giommi_2015} Giommi P., 2015, JHEAp, 7, 173
\bibitem[\protect\citeauthoryear{Giommi et al.}{2018}]{openuniverse} Giommi P., et al., 2018, arXiv, arXiv:1805.08505
\bibitem[\protect\citeauthoryear{Giommi et al.}{2019a}]{Giommi_2019a} 
Giommi P., et al., 2019a, A\&A, 631, A116 
\bibitem[\protect\citeauthoryear{Giommi et al.}{2019b}]{Giommi_2019b} 
Giommi P., et al., 2019b, in preparation
\bibitem[\protect\citeauthoryear{Gregory \& Condon}{1991}]{Gregory_1991}
Gregory P.~C., Condon J.~J., 1991, ApJS, 75, 1011
\bibitem[\protect\citeauthoryear{Halzen \& Zas}{1997}]{halzen97} {Halzen}
 F., {Zas} E., 1997, ApJ, 488, 669
\bibitem[\protect\citeauthoryear{Huber for the IceCube Collaboration}{2019}]{Huber:2019lrm} M.~Huber [IceCube Collaboration],
  arXiv:1908.08458 [astro-ph.HE].
\bibitem[\protect\citeauthoryear{IceCube Collaboration}{2013}]{ICECube13}
  IceCube Collaboration, 2013, Science, 342, 1242856
\bibitem[\protect\citeauthoryear{IceCube Collaboration}{2018}]{datarelease} 
  IceCube Collaboration, 2018, DOI:10.21234/B4KS6S
\bibitem[\protect\citeauthoryear{IceCube Collaboration}{2014}]{ICECube14}
  IceCube Collaboration, 2014, Phys. Rev. Lett., 113, 101101
\bibitem[\protect\citeauthoryear{IceCube
    Collaboration}{2015a}]{ICECube15_1} IceCube Collaboration, 2015a,
  Contributions to the 34th International Cosmic Ray Conference (ICRC
  2015), p. 45 (arXiv:1510.05223)
\bibitem[\protect\citeauthoryear{IceCube
    Collaboration}{2015b}]{ICECube15_2} IceCube Collaboration, 2015b,
  Contributions to the 34th International Cosmic Ray Conference (ICRC
  2015), p. 37 (arXiv:1510.05223)
\bibitem[\protect\citeauthoryear{IceCube
    Collaboration}{2017a}]{ICECube17_1} IceCube Collaboration, 2017a,
  Contributions to the 35th International Cosmic Ray Conference (ICRC
  2017), p. 30 (arXiv:1710.01191)
\bibitem[\protect\citeauthoryear{IceCube
    Collaboration}{2017b}]{ICECube17_2} IceCube Collaboration, 2017b,
  Contributions to the 35th International Cosmic Ray Conference (ICRC
  2017), p. 54 (arXiv:1710.01191)
\bibitem[\protect\citeauthoryear{IceCube
    Collaboration}{2017c}]{ICECube17_3} IceCube Collaboration, 2017c,
  Contributions to the 35th International Cosmic Ray Conference (ICRC
  2017), p. 31 (arXiv:1710.01179)
 \bibitem[\protect\citeauthoryear{IceCube
    Collaboration}{2017d}]{threeneutrinos} IceCube Collaboration, 2017d,
    A\&A 607, 115I
\bibitem[\protect\citeauthoryear{IceCube Collaboration et al.}{2018a}]{icfermi}
 IceCube Collaboration, 2018, Science, 361, 147
\bibitem[\protect\citeauthoryear{IceCube Collaboration}{2018b}]{iconly}
 IceCube Collaboration et al., 2018, Science, 361, eaat1378
\bibitem[\protect\citeauthoryear{Lucarelli et al.}{2017}]{Lucarelli_2017}
Lucarelli F., et al., 2017, ApJ, 846, 121
\bibitem[\protect\citeauthoryear{Lucarelli et al.}{2019}]{Lucarelli_2019}
Lucarelli F., et al., 2019, ApJ, 870, 136
\bibitem[\protect\citeauthoryear{Mannheim}{1995}]{mannheim95} Mannheim K., 1995, Astroparticle Physics, 3, 295
\bibitem[\protect\citeauthoryear{Massaro et al.}{2015}]{5bzcat}
Massaro E., et al. 2015, Ap\&SS 357, 75
\bibitem[\protect\citeauthoryear{M{\"u}cke et al.}{2003}]{mueckeetal03}
  {M{\"u}cke} A. et al., 2003, Astroparticle Physics, 18, 593
\bibitem[\protect\citeauthoryear{Padovani \& Giommi}{1995}]{padgio95}
  Padovani P., Giommi P., 1995, ApJ, 444, 567
\bibitem[\protect\citeauthoryear{Padovani \& Resconi}{2014}]{Pad_2014}
  Padovani P., Resconi E., 2014, MNRAS, 443, 474 
\bibitem[\protect\citeauthoryear{Padovani \& Giommi}{2015}]{padovanibsv1}
  Padovani P., Giommi P., 2015, MNRAS, 446, L41
\bibitem[\protect\citeauthoryear{Padovani et al.}{2016}]{Padovani_2016}
  Padovani P., Resconi E., Giommi P., Arsioli B., Chang Y.~L., 2016, MNRAS,
  457, 3582
\bibitem[\protect\citeauthoryear{Padovani et al.}{2017}]{Padovani_2017}
  Padovani P., et al., 2017, A\&ARv, 25, 2
\bibitem[\protect\citeauthoryear{Padovani, Turcati \& Resconi}{2018}]{outflows}
  Padovani P., Turcati A., \& Resconi E., 2018, MNRAS, 477, 3469
\bibitem[\protect\citeauthoryear{Padovani et 
al.}{2018}]{dissecting} Padovani P., Giommi 
P., Resconi E., Glauch T., Arsioli B.,  
Sahakyan N., Huber M., 2018, MNRAS, 480, 192
\bibitem[\protect\citeauthoryear{Padovani et 
al.}{2019}]{Padovani_2019} Padovani P., Oikonomou F., Petropoulou M., Giommi 
P., Resconi E., 2019, MNRAS, 484, L104
\bibitem[\protect\citeauthoryear{Paiano et al.}{2019}]{Paiano_2019} Paiano S., Padovani P., 
Falomo R., Giommmi P., Scarpa R., Treves A. 2019, The Astronomer's Telegram, 13202
\bibitem[\protect\citeauthoryear{Paliya et al.}{2019}]{Paliya}
Paliya, V. S., Dom\'inguez, A., Ajello, M., Franckowiak, A., Hartmann, D., 2019, ApJ, 882, L3
\bibitem[\protect\citeauthoryear{Palladino \& Vissani}{2017}]{Palladino_2017} Palladino A., Vissani F., 2017, A\&A, 604, A18 
\bibitem[\protect\citeauthoryear{Plante, et al.}{2008}]{conesearch} Plante R., Williams R., Hanisch R., Szalay A., 2008, ivoa.spec
\bibitem[\protect\citeauthoryear{Resconi et al.}{2017}]{Resconi_2017}
  Resconi E., Coenders S., Padovani P., Giommi P., Caccianiga L., 2017,
  MNRAS, 468, 597
\bibitem[\protect\citeauthoryear{Schneider}{2019}]{2019ICRC...36.1004S} 
Schneider A., on behalf of the IceCube Collaboration, 2019, 36th 
International Cosmic Ray Conference, 1004  
\bibitem[\protect\citeauthoryear{Snios et al.}{2019}]{Snios} Snios B., et al., 2019, ApJ, 879, 8
\bibitem[\protect\citeauthoryear{Stettner}{2019}]{2019ICRC...36.1017S} 
Stettner J., on behalf of the IceCube Collaboration, 2019, 36th 
International Cosmic Ray Conference, 1017
\bibitem[\protect\citeauthoryear{Stickel et al.}{1991}]{Stickel_1991}
  Stickel M., Padovani P., Urry C. M., Fried J. W., K\"uhr H., 1991, ApJ,
  374, 431
\bibitem[\protect\citeauthoryear{Stocke et al.}{1991}]{Stocke_1991} Stocke
  J.~T., Morris S.~L., Gioia I.~M., Maccacaro T., Schild R., Wolter A.,
  Fleming T.~A., Henry J.~P., 1991, ApJS, 76, 813
\bibitem[\protect\citeauthoryear{Tody, et al.}{2012}]{specsearch} Tody D., et al., 2012, ivoa.spec
\bibitem[\protect\citeauthoryear{Urry \& Padovani}{1995}]{UP95} Urry C.~M.,
  Padovani P., 1995, PASP, 107, 803
\bibitem[\protect\citeauthoryear{Voges et al.}{1999}]{RASS_1999} Voges W., et al., 1999, A\&A, 349, 389  
\bibitem[\protect\citeauthoryear{Voges et al.}{2000}]{RASS_2000} Voges W., et al., 2000, IAUC, 7432, 1
\bibitem[\protect\citeauthoryear{Walker et al.}{2018}]{Walker_2018} Walker R.~C., Hardee P.~E., 
Davies F.~B., Ly C., Junor W., 2018, ApJ, 855, 128
\end{thebibliography}
\end{document}